\pdfoutput=1
\documentclass[11pt,letterpaper]{article}

\usepackage[utf8]{inputenc}
\usepackage[T1]{fontenc}
\usepackage[sc]{mathpazo}
\usepackage[margin=1in,headheight=22pt]{geometry}
\usepackage{microtype}
\usepackage{graphicx}
\usepackage{amsmath,amssymb,bm}
\usepackage{booktabs}
\usepackage{xcolor}
\usepackage[font=small,labelfont=bf,labelsep=period]{caption}
\usepackage{fancyhdr}
\usepackage{titlesec}
\usepackage[round,authoryear]{natbib}

\definecolor{linkblue}{RGB}{24,78,136}
\definecolor{rulegray}{RGB}{180,180,180}

\usepackage[
  colorlinks=true,
  linkcolor=linkblue,
  citecolor=linkblue,
  urlcolor=linkblue,
  pdfauthor={Pu Yang, Yuxin Liu, Zhengyang Fang, Bin Zhang, Deshan Feng},
  pdftitle={Full-waveform inversion with adaptive anisotropic Gaussian primitives}
]{hyperref}

\titleformat{\section}{\large\bfseries}{\thesection}{0.75em}{}
\titleformat{\subsection}{\normalsize\bfseries}{\thesubsection}{0.75em}{}
\titleformat{\subsubsection}{\normalsize\itshape}{\thesubsubsection}{0.75em}{}

\renewcommand{\headrulewidth}{0.4pt}
\renewcommand{\headrule}{\hbox to\headwidth{\color{rulegray}\leaders\hrule height \headrulewidth\hfill}}

\fancypagestyle{plain}{%
  \fancyhf{}
  \fancyfoot[C]{\footnotesize\thepage}
  \renewcommand{\headrulewidth}{0pt}
  
}

\begin{document}

\thispagestyle{plain}

% Title block (arXiv preprint style)
\begin{center}
  \rule{\linewidth}{1.6pt}\\[0.9em]
  {\LARGE\bfseries Full-waveform inversion with adaptive anisotropic Gaussian primitives\par}
  \vspace{0.6em}
  \rule{\linewidth}{0.6pt}\\[1.2em]
  {\normalsize
    \textbf{Pu Yang}, \textbf{Yuxin Liu}, \textbf{Zhengyang Fang}, \textbf{Bin Zhang}, and \textbf{Deshan Feng}\par}
  \vspace{0.5em}
  {\small\itshape
    School of Geosciences and Info-physics, Central South University, Changsha 410083, China\par}
  \vspace{0.8em}
\end{center}

\begin{abstract}
\noindent
Reparameterizing the velocity model has emerged as an effective way to regularize full-waveform inversion (FWI) and reduce its sensitivity to the starting model. Existing parameterizations, however, either fix the position and shape of every basis function in advance, as in grids and splines, or control the spatial structure of their updates only indirectly through the shared weights of a neural network. We propose Gauss-FWI, which represents the velocity perturbation as a sum of anisotropic Gaussian primitives with learnable amplitudes, centres, scales, and orientations. The basis itself thus adapts to the subsurface. Primitives move onto interfaces, stretch along layers, and rotate with the dip, so model updates follow the geological structure without a hand-designed regularizer. A gradient-energy criterion adds finer zero-amplitude primitives where the data call for more detail, increasing resolution progressively without perturbing the current model. All parameters receive analytic gradients through the adjoint-state method on a fixed simulation grid, so the representation fits into existing FWI workflows and combines directly with frequency continuation. Each primitive keeps an explicit position, extent, and orientation, and the representation Jacobian links the waveform gradient to the resulting model update, which makes the behaviour of the inversion interpretable. In two-dimensional acoustic examples, Gauss-FWI uses about fifty times fewer parameters than implicit FWI and recovers the Marmousi, Overthrust, and BP salt models more accurately than grid, multiscale, B-spline, and implicit parameterizations. It remains accurate under missing low frequencies, noise, and discretization mismatch, and it recovers the main structure from starting models where the other methods fail, such as a linear model without the BP salt body.

\vspace{0.6em}
\end{abstract}

\vspace{0.5em}

\section{Introduction}
Full-waveform inversion (FWI) is an important tool for high-resolution subsurface characterization in seismic exploration, where accurate velocity models support the imaging of complex geological structures and the assessment of subsurface resources \citep{virieux2009}. By iteratively matching simulated wavefields to recorded seismic data, FWI exploits phase and amplitude information to estimate subsurface physical properties \citep{lailly1983,tarantola1984,pratt1998}. The nonlinear waveform objective is nevertheless sensitive to the initial model and the available frequency content. In particular, when predicted and observed arrivals differ sufficiently in phase, a least-squares waveform objective may match the wrong cycles. This cycle-skipping problem can drive local optimization towards an erroneous model, especially when the starting velocity is inaccurate or low-frequency data are unavailable. Errors in the long-wavelength velocity structure can then compromise the recovery of finer features.

To mitigate cycle skipping, existing approaches modify the data, model space or misfit function to guide the inversion towards a more accurate velocity model. Data-domain multiscale methods simplify early phase matching either by frequency continuation, which starts from low-frequency data and progressively incorporates higher frequencies \citep{bunks1995,fichtner2013}, or by decomposing observed and predicted seismograms across wavelet scales \citep{yuan2014}. Model-domain multiscale methods instead restrict early updates to a coarse representation and progressively introduce finer degrees of freedom. \citet{barnier2019} use a B-spline basis with refined control grids. Model-extension methods introduce additional degrees of freedom to relax the wavefield-matching problem before imposing consistency with a physical model \citep{barnier2023}. Other methods replace the conventional least-squares objective with measures that account for waveform shifts \citep{bozdag2011,wu2014}. Traveltime-based objectives compare the timing of corresponding arrivals \citep{hu2026}, whereas optimal-transport objectives measure the cost of rearranging transformed seismic signals and can provide a more favourable misfit landscape for shifted waveforms \citep{yang2018}. Together, these methods illustrate how the choice of data representation, model parameterization and misfit influences the inversion path.

Among these approaches, model-domain multiscale methods suggest that the way a velocity model is represented can itself influence the course of inversion. A coarse parameterization initially restricts the available updates to larger spatial scales, while refinement allows smaller-scale variations to be introduced. More generally, even with the same data and waveform objective, different representations map a physical-model gradient into different spatial updates. Basis functions can couple neighbouring locations and favour particular scales, thereby influencing the optimization path \citep{loris2007,kadu2017}. Such ideas predate neural representations. \citet{sun2003} investigated different basis functions for acoustic waveform inversion, and spline representations allow non-uniform control-point spacing and progressive refinement \citep{barnier2019}. These observations motivate the study of model representations that can adapt their spatial constraints during inversion \citep{bodin2009}.

Neural reparameterization extends these model-space approaches by using a neural network to represent the subsurface properties \citep{wu2019cnn,he2021,zhu2022}. In implicit full-waveform inversion (IFWI), a coordinate-based network maps spatial coordinates to a continuous property field, and its parameters are optimized through the waveform objective \citep{sun2023,zhang2023,du2024,wang2026}. Because the network parameters are shared across spatial locations, their updates can couple changes at different points in the model. This coupling depends on the network architecture and can influence the relative rates at which model structures of different scales are recovered \citep{rahaman2019,chen2026wave,wu2025}. In the examples studied by \citet{chen2026wave}, an implicit representation favours large-scale recovery and improves tolerance to inaccurate starting models, but its fine-scale features converge more slowly. This trade-off has motivated the design of neural representations with different scale-dependent biases.

Recent work has introduced additional inductive biases to adjust this balance \citep{chen2025,kang2025,chen2026pert}. Multi-grid parametric encoding (MPE) combines learnable features on grids at multiple resolutions \citep{muller2022,audia2025}; in FWI, \citet{chen2026wave} show that these localized features can accelerate the convergence of fine-scale components when starting from a smooth initial model. The INR-grid (IG) representation combines such grid features with features generated by an implicit neural representation, seeking a balance between local detail and the smoother structure provided by the coordinate network \citep{chen2026wave}. Other neural parameterizations use architectural choices such as convolution, interpolation, and upsampling to constrain spatial relationships in the reconstructed model \citep{wu2019cnn,zhu2022}. In particular, \citet{wu2025} examine how interpolation in deep decoder architectures introduces a model-domain multiscale bias. The benefits of these representations depend on the inversion setting. For example, \citet{chen2026wave} report that although MPE improves fine-scale convergence from a smooth starting model, it can be less effective when the starting model is inaccurate. This dependence makes the choice of a suitable representation and its scale-dependent bias an important consideration for each inversion problem.

The dependence on representation raises a further question for model-space design. Neural parameterizations can induce useful spatial correlations, but the location, scale, and orientation of their updates are generally controlled indirectly through network parameters. This makes these properties difficult to adjust and interpret individually as the inversion evolves. Gaussian basis functions have previously been used in waveform inversion to parameterize velocity models with fixed basis centres and widths \citep{borcea2023} or to estimate smooth starting models via sparse Gaussian radial basis functions \citep{chen2020swi}. Rather than keeping the basis geometry fixed, we draw inspiration from Gaussian splatting, originally developed for radiance-field rendering, which represents a scene with anisotropic Gaussian primitives whose geometry and spatial density are adapted during optimization \citep{kerbl2023}. Learning the anisotropic geometry of the primitives and adaptively adding finer ones during inversion provides a direct way to adjust the spatial structure and local capacity of a model representation in FWI.

Building on this idea, we propose Gauss-FWI, which represents the velocity perturbation as a superposition of anisotropic Gaussian primitives. Each primitive has an amplitude, a centre, principal scales, and an orientation. These parameters are optimized against the waveform objective, so the positions and shapes of the local basis functions can evolve with the inversion while the wave equation remains discretized on a fixed computational grid. The Gaussian derivatives with respect to amplitude and geometry are available analytically and connect the physical-model gradient to parameter updates. To introduce finer-scale capacity, a gradient-energy criterion selects regions for refinement. Existing primitives are retained, and finer primitives are introduced with zero amplitude. This construction preserves the current velocity field at the refinement event while providing additional local update directions for subsequent iterations.

The explicit parameterization also provides a basis for analysing the resulting model updates. We examine the local update modes associated with amplitude and geometry, and relate the representation Jacobian to a model-space update operator under a local, frozen-metric approximation. This analysis distinguishes changes in the geometry of existing update modes from the addition of new modes through refinement. Field preservation follows from the zero-amplitude additive construction; it does not by itself guarantee improved conditioning or recovery. The local operator is likewise an analysis tool rather than an exact description of the complete trajectory of an optimizer with momentum, weight decay, and geometric projections.

We evaluate Gauss-FWI using two-dimensional (2-D) constant-density acoustic examples and compare it with grid-based, B-spline, and neural parameterizations. Model-recovery comparisons, geometry and refinement ablations, and tests with different starting models and data conditions are used to assess both the benefits and the limitations of the representation. The experiments address how explicit geometry and adaptive capacity affect the recovery of the tested velocity models, rather than establishing a general guarantee against cycle skipping.

\section{Method}

Gauss-FWI writes the velocity model as a sum of Gaussian ellipses that the inversion can move, stretch, and rotate, and it adds smaller ellipses where finer structure is needed (Fig.~\ref{fig:overview}). The wave equation is solved on a fixed grid throughout.

\subsection{Gaussian representation}
\label{sec:gaussian-rep}

The velocity at position $\mathbf{x} = (z, x)^\top$ is the initial model $v_0$ plus $K$ anisotropic Gaussian primitives:
\begin{equation}
v_\theta(\mathbf{x}) = v_0(\mathbf{x}) + s \sum_{k=1}^K a_k g_k(\mathbf{x}), \qquad
g_k(\mathbf{x}) = \exp\!\left( -\frac{1}{2} (\mathbf{x} - \boldsymbol{\mu}_k)^\top \boldsymbol{\Sigma}_k^{-1} (\mathbf{x} - \boldsymbol{\mu}_k) \right),
\label{eq:velocity-def}
\end{equation}
where $s$ is a fixed velocity scale. The covariance $\boldsymbol{\Sigma}_k = \mathbf{R}(\theta_k)^\top \operatorname{diag}(s_{k,0}^2, s_{k,1}^2) \mathbf{R}(\theta_k)$ is defined by two principal standard deviations and a rotation by the angle $\theta_k$. Each primitive thus has six parameters: an amplitude $a_k$, a centre $\boldsymbol{\mu}_k$, two log-scales $\ell_{k,j} = \ln s_{k,j}$ and an orientation $\theta_k$. Lengths are measured in grid cells.

\begin{figure}[tbp]
  \centering
  \includegraphics[width=\textwidth]{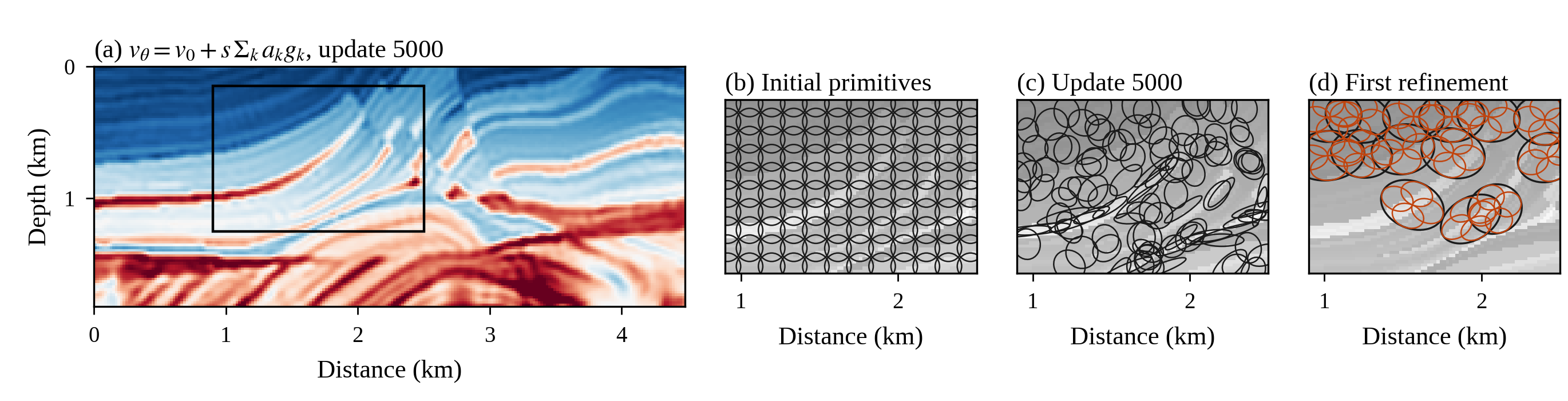}
  \caption{Overview of Gauss-FWI on the Marmousi model. (a) The recovered velocity, the starting model plus a sum of anisotropic Gaussian primitives, after 5000 updates; the box marks the window of (b)--(d). (b) One level of the initial dictionary (one-standard-deviation contours over the true model). (c) The same primitives after the inversion: they have moved, stretched, and rotated along the layers. (d) The first refinement (update 200): refined parents (black) keep their parameters and receive four zero-amplitude children (orange) with half their scales, so the velocity model is unchanged.}
  \label{fig:overview}
\end{figure}

The primitives are initialized on $L$ dyadic levels. Level $l$ holds $2^l \times 2^{l+1}$ isotropic primitives on a uniform lattice, which gives $K_0 = 682$ primitives for $L = 5$ (Fig.~\ref{fig:overview}(b)). All amplitudes start at zero, so the inversion starts exactly from $v_0$.

The model $\mathbf{v}_\theta$ is evaluated on the grid of $N$ nodes used by the wave-equation solver, and the inversion minimizes the normalized least-squares objective,
\begin{equation}
\mathcal{J}(\theta) = \frac{\|\mathcal{F}(\mathbf{v}_\theta) - \mathbf{d}_{\text{obs}}\|_2^2}{\|\mathbf{d}_{\text{obs}}\|_2^2},
\label{eq:misfit-def}
\end{equation}
where $\mathcal{F}$ is the forward-modelling operator and $\mathbf{d}_{\text{obs}}$ is the observed data. The adjoint-state method \citep{tarantola1984,plessix2006} gives the gradient with respect to the gridded velocity, $\mathbf{g}_v = \partial \mathcal{J} / \partial \mathbf{v}$, and the chain rule maps it to the Gaussian parameters:
\begin{equation}
\mathbf{g}_\theta = \mathbf{B}_\theta^\top \mathbf{g}_v, \qquad \mathbf{B}_\theta = \frac{\partial \mathbf{v}_\theta}{\partial \theta}.
\label{eq:parameter-gradient}
\end{equation}

\subsection{Learning the geometry}
\label{sec:update-modes}

\begin{figure}[tbp]
  \centering
  \includegraphics[width=0.6\textwidth]{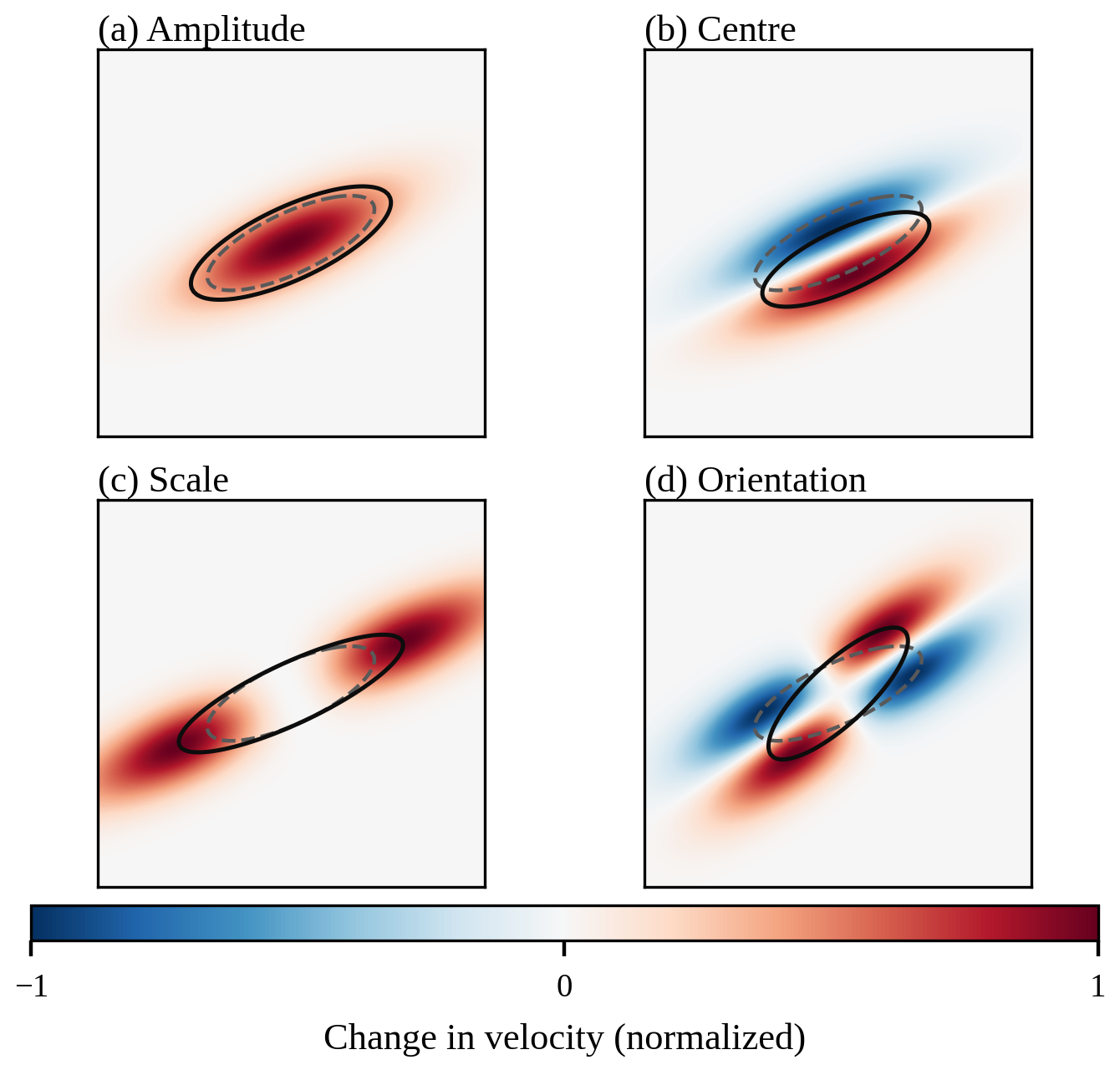}
  \caption{Change of the velocity model when one kind of parameter of a primitive is perturbed: (a) amplitude, (b) centre, moved across the major axis, (c) scale, enlarged along the major axis, and (d) orientation. Colour: velocity after minus before, normalized to unit maximum. Dashed and solid curves: half-maximum contour before and after the perturbation.}
  \label{fig:modes}
\end{figure}

Conventional basis functions, such as grid cells or B-splines, have prescribed shapes and positions, and the inversion only determines their coefficients. In Gauss-FWI all six parameters of each primitive are updated by the waveform gradient. Each column of $\mathbf{B}_\theta$ is the pattern by which one parameter changes the model (Fig.~\ref{fig:modes}). With $(u_{k,0}, u_{k,1})^\top = \mathbf{R}(\theta_k)(\mathbf{x} - \boldsymbol{\mu}_k)$ the coordinates along the principal axes, these patterns are
\begin{align}
\frac{\partial v_\theta}{\partial a_k} &= s\, g_k, &
\frac{\partial v_\theta}{\partial \boldsymbol{\mu}_k} &= s\, a_k\, g_k\, \boldsymbol{\Sigma}_k^{-1} (\mathbf{x} - \boldsymbol{\mu}_k), \label{eq:mode-amp} \\
\frac{\partial v_\theta}{\partial \ell_{k,j}} &= s\, a_k\, g_k\, \frac{u_{k,j}^2}{s_{k,j}^2}, &
\frac{\partial v_\theta}{\partial \theta_k} &= s\, a_k\, g_k\, u_{k,0} u_{k,1} \left( \frac{1}{s_{k,1}^2} - \frac{1}{s_{k,0}^2} \right). \label{eq:mode-angle}
\end{align}

The amplitude raises or lowers the velocity under the primitive, the centre translates it, the scales widen or narrow it along each axis, and the orientation rotates it. The geometric parameters therefore change the spatial shape of the update itself, so that primitives can move onto interfaces, elongate along layers and turn with the dip (Fig.~\ref{fig:overview}(c)). The geometric derivatives are proportional to $a_k$: a primitive with zero amplitude receives only an amplitude update, and its geometry starts to adapt once it contributes to the model.

\subsection{Refinement}
\label{sec:refinement}

Adapting the geometry rearranges the existing primitives but does not increase their number. To add resolution where it is needed, the squared gradient is accumulated on the grid between refinement events, $E(\mathbf{x}) \leftarrow E(\mathbf{x}) + |g_v(\mathbf{x})|^2$, and each primitive is scored by the energy under its footprint:
\begin{equation}
S_k = \sum_{\mathbf{x}} g_k(\mathbf{x}) E(\mathbf{x}).
\label{eq:parent-score}
\end{equation}

At each event, the fraction $\alpha$ of primitives with the highest scores is refined, up to a total of $K_{\max}$ primitives. A primitive is refined at most once, and only if its halved scales remain above the minimum scale $s_{\min}$.

A refined primitive is kept unchanged and receives four children with half its scales, the same orientation and zero amplitude, placed at $\boldsymbol{\mu}_p \pm \frac{1}{2} s_{p,j} \mathbf{e}_{p,j}$ along its two principal axes $\mathbf{e}_{p,j}$ (Fig.~\ref{fig:overview}(d)). Because the children have zero amplitude, the velocity model, the simulated data and the objective are identical before and after refinement. Refinement only adds finer primitives that the subsequent updates can use.

\subsection{Constraints and optimization}
\label{sec:optimization}

After every update, the centres are kept at least $c_{\text{margin}}$ cells inside the grid and the principal standard deviations within $[s_{\min}, s_{\max}]$, where $s_{\min}$ prevents primitives narrower than the grid can sample. The vertical extent of each primitive is also bounded,
\begin{equation}
\sigma_{z, k} = \sqrt{s_{k,0}^2 \cos^2\theta_k + s_{k,1}^2 \sin^2\theta_k} \le \sigma_{z, \max},
\label{eq:vertical-std}
\end{equation}
so that a single primitive cannot couple shallow and deep updates, and an optional bound $r_{\max}$ limits its aspect ratio.

The four types of parameters are optimized with AdamW using separate learning rates, $\eta_a = 0.003$, $\eta_\mu = 0.05$~cells, $\eta_\ell = 0.002$ and $\eta_\theta = 0.005$~rad, which decay exponentially to a factor $\gamma_{\text{lr}}$ of their initial values over the $T$ updates. A weight decay of 0.01 is applied to the amplitudes only. All experiments use $s_{\min} = 1.5$ and $s_{\max} = 150$~cells, $\alpha = 0.25$ and $K_{\max} = 4096$; the remaining settings are given with each experiment.

\section{Numerical experiments}

\subsection{Test models and experimental setup}

\begin{figure}[tbp]
  \centering
  \includegraphics[width=\textwidth]{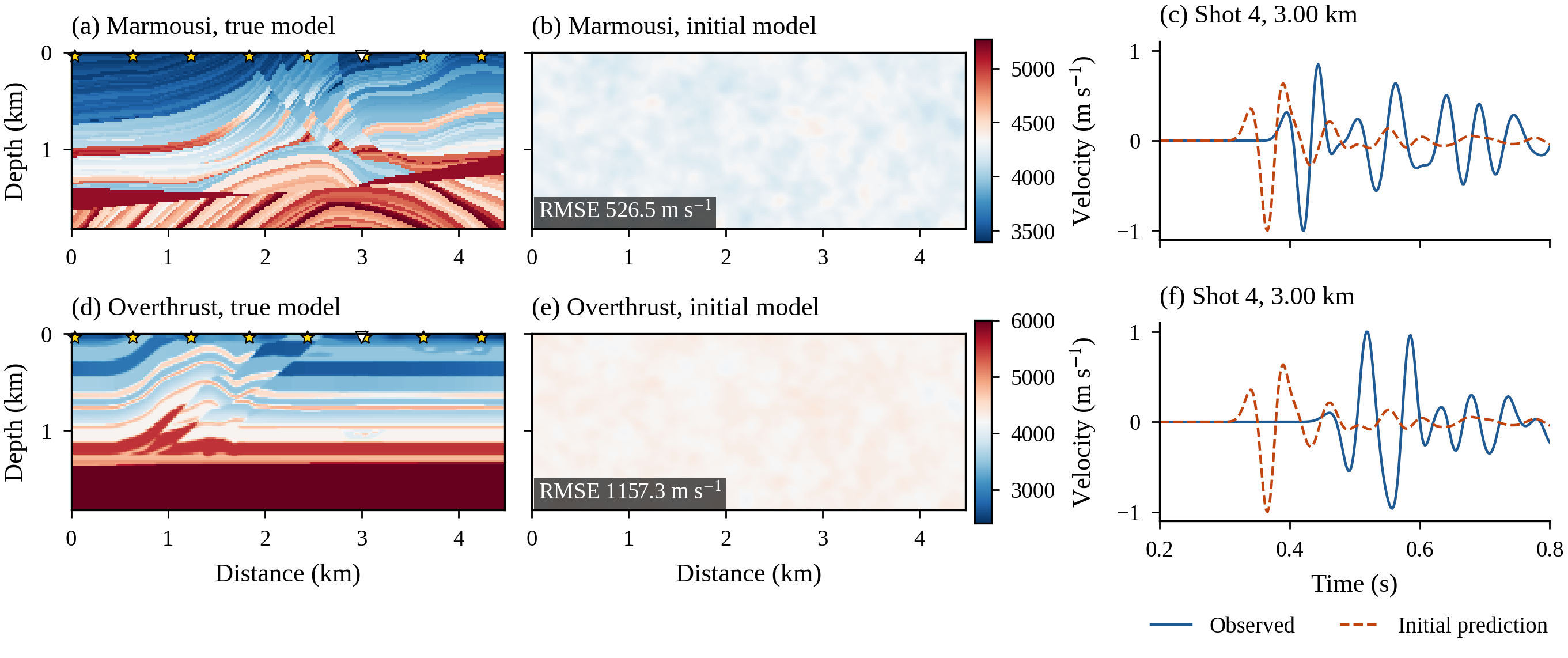}
  \caption{Test models, acquisition and the common starting model. Upper row: Marmousi-derived model. Lower row: Overthrust-derived model. (a), (d) True P-wave velocity models with the eight sources (stars) and the receiver used for the displayed traces (triangle). (b), (e) The common starting model shown on the respective velocity colour scales. (c), (f) Synthetic observations and initial predictions for the fourth shot at the receiver at a horizontal distance of 3.00~km. Each trace is normalized separately over the displayed time window.}
  \label{fig:initial-waveform}
\end{figure}

In this section, we use 2-D P-wave velocity models derived from Marmousi \citep{versteeg1994} and Overthrust (Fig.~\ref{fig:initial-waveform}(a) and (d), respectively). Both models are defined on a grid of $225 \times 92$ points ($N_x \times N_z$) with a uniform spacing of 20~m, spanning 4.48~km horizontally and 1.82~km in depth.

Both models use the same acquisition geometry: eight shots spaced 600~m apart and 224 receivers per shot spaced 20~m apart. The sources and receivers are located 40~m below the top of the model. We use a Ricker source wavelet with a peak frequency of 15~Hz and a peak time of 0.1~s. Each receiver trace contains 1500 samples at 2~ms intervals. In the baseline experiments, we generate synthetic observations from the true models and compute predicted data during inversion using Deepwave's 2-D constant-density acoustic propagator \citep{richardson2026}. It uses fourth-order spatial and second-order temporal finite differences, with a zero-pressure boundary at the top of the model and 20-cell perfectly matched layers at the lateral and bottom boundaries.

We use a common starting model for the Marmousi inversions and as one of the starting models for Overthrust (Fig.~\ref{fig:initial-waveform}(b) and (e)). We obtain it by scaling the output of a randomly initialized coordinate-based neural network using the mean and standard deviation of a vertical velocity profile extracted from the true Marmousi model.

Each inversion runs for 5000 updates unless otherwise stated. The baseline Gauss-FWI configuration uses a velocity scaling factor of $s = 500$~m/s. When geometry is adaptive, the vertical standard deviation is limited to $\sigma_{z,\max} = 160$~m, the centres are kept at least $c_{\text{margin}} = 1$~cell inside the grid, and the aspect ratio is capped at $r_{\max} = 40$. When refinement is enabled, it is performed every 200 updates and the accumulated gradient energy is reset after each event.

\subsection{Effects of geometry adaptation and refinement}
\label{sec:ablation}

\begin{figure}[tbp]
  \centering
  \includegraphics[width=\textwidth]{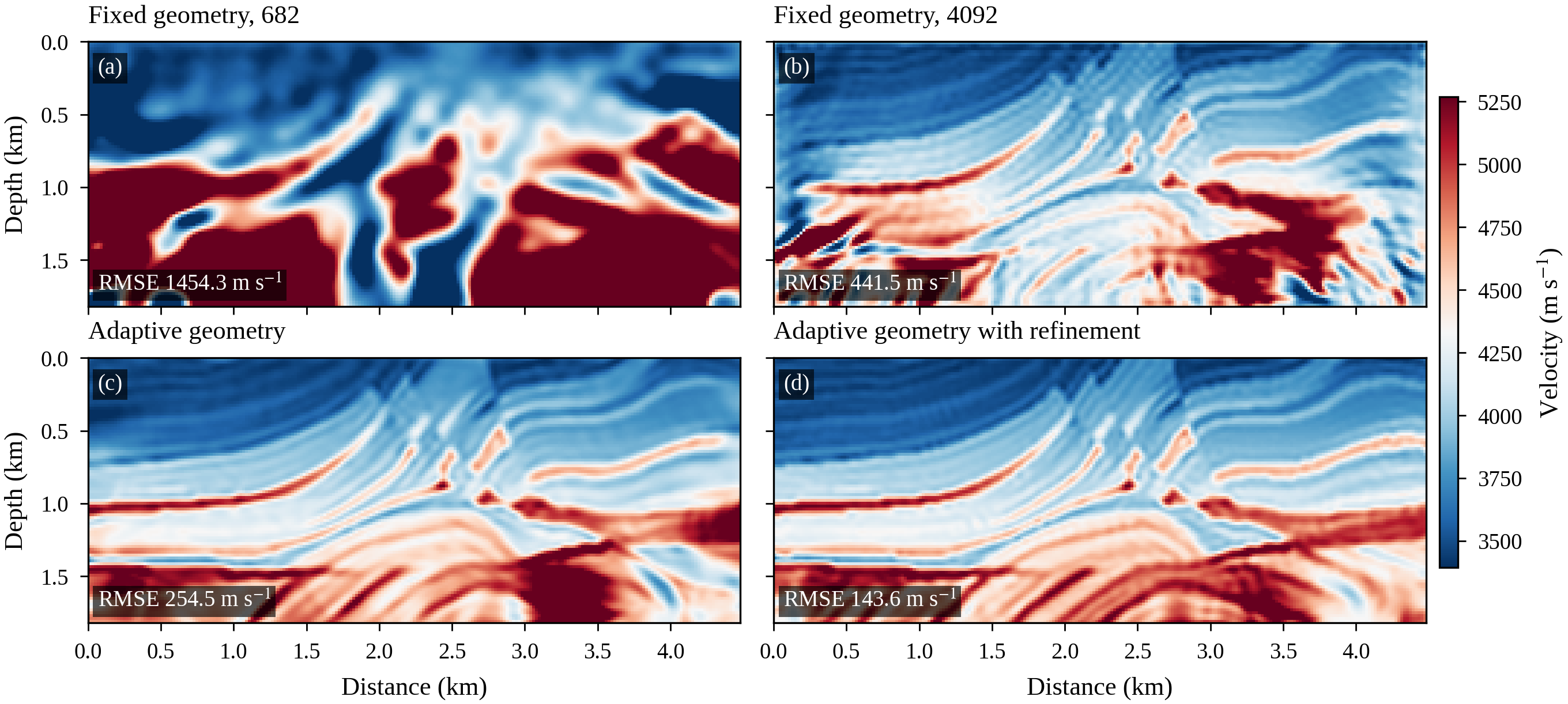}
  \caption{Effect of geometry adaptation and refinement on the Marmousi model. (a) 682 dyadic primitives with fixed geometry, amplitudes only. (b) 4092 primitives with fixed geometry on a uniform $44 \times 93$ lattice, amplitudes only. (c) 682 dyadic primitives with adaptive geometry and no refinement. (d) Adaptive geometry with refinement (Gauss-FWI). All inversions run for 5000 updates. The RMSE of each recovered model is shown in its panel.}
  \label{fig:marmousi-geometry}
\end{figure}

We run four Marmousi inversions to assess how the initial Gaussian dictionary, geometry adaptation, and refinement affect velocity recovery. Each inversion starts from the velocity model in Fig.~\ref{fig:initial-waveform}(b), uses the same synthetic observations, and runs for 5000 updates. The AdamW settings described in Section~\ref{sec:optimization} are used for all four inversions, with a final learning-rate factor of $\gamma_{\text{lr}} = 0.95$. They differ only in which parameters are trained and whether refinement is active. The fixed-geometry representation contains the 682 dyadic primitives and trains their amplitudes only. The dense fixed-geometry representation places 4092 primitives on a uniform $44 \times 93$ ($N_z \times N_x$) lattice and also trains amplitudes only. The adaptive-geometry representation starts from the 682 dyadic primitives and trains all six parameters of each, without refinement. The fourth representation is the full Gauss-FWI, which in addition refines every 200 updates. Fig.~\ref{fig:marmousi-geometry} shows the recovered models after 5000 updates, with the root-mean-square error (RMSE) measured against the true model.

With 682 fixed primitives, the recovered model contains large, blocky velocity errors and has an RMSE of 1454.3~m/s (Fig.~\ref{fig:marmousi-geometry}(a)). Increasing the number of fixed primitives to 4092 reduces the RMSE to 441.5~m/s and recovers most of the shallow layering (Fig.~\ref{fig:marmousi-geometry}(b)), but large high-velocity anomalies remain below about 1.2~km depth, at the left boundary and between 2.5 and 4~km distance.

The adaptive representation in Fig.~\ref{fig:marmousi-geometry}(c) uses the same 682 primitives as Fig.~\ref{fig:marmousi-geometry}(a) but also updates their centres, scales and orientations. The reflectors become continuous along the layers, and the RMSE decreases from 1454.3 to 254.5~m/s. Figs~\ref{fig:marmousi-geometry}(b) and (c) have the same number of trainable parameters, 4092, yet the adaptive representation has a 42\% lower RMSE. In the dense fixed representation, every parameter is the weight of a prescribed pattern. In the adaptive representation, five of the six parameters of each primitive move, resize, or rotate the pattern itself, so the spatial patterns available for model updates evolve during the inversion. For this Marmousi configuration, spending parameters on geometry is more effective than spending them on more fixed primitives. Nevertheless, a high-velocity overshoot remains in the deep part of Fig.~\ref{fig:marmousi-geometry}(c), between about 3 and 3.7~km distance.

Refinement adds finer primitives where the accumulated gradient energy is high. Starting from 682 primitives, the full method reaches 3358 primitives by update 4600, sharpens the thin layers, and reduces the deep overshoot (Fig.~\ref{fig:marmousi-geometry}(d)). Its RMSE of 143.6~m/s is 44\% lower than that of the adaptive representation without refinement.

\subsection{Update structure}
\label{sec:update-structure}

The adaptive-geometry representation recovers the Marmousi layers more clearly than the dense fixed-geometry representation at the same budget of 4092 trainable parameters (Fig.~\ref{fig:marmousi-geometry}(b) and (c)). To examine how the two representations use these parameters, we compare the velocity responses they produce for the same gradient. To first order, the Jacobian $\mathbf{B}_\theta$ maps a parameter increment $\Delta\theta$ to the velocity change $\mathbf{B}_\theta\Delta\theta$. Holding the diagonal scaling $\mathbf{D}_\theta$ from AdamW fixed, we write the response to the current waveform gradient $\mathbf{g}_v$ as
\begin{equation}
\Delta\mathbf{v}_{\mathrm{curr}} = -\mathbf{B}_\theta \mathbf{D}_\theta \mathbf{B}_\theta^{\top} \mathbf{g}_v = -\mathbf{P}_\theta \mathbf{g}_v.
\label{eq:update-operator}
\end{equation}
The columns of $\mathbf{B}_\theta$ give the spatial patterns available to the inversion for small parameter changes. With fixed geometry, these patterns do not change during inversion and $\mathbf{P}_\theta$ varies only through $\mathbf{D}_\theta$. With adaptive geometry, the patterns themselves can change as the primitives move, stretch, and rotate.

\begin{figure}[tbp]
  \centering
  \includegraphics[width=\textwidth]{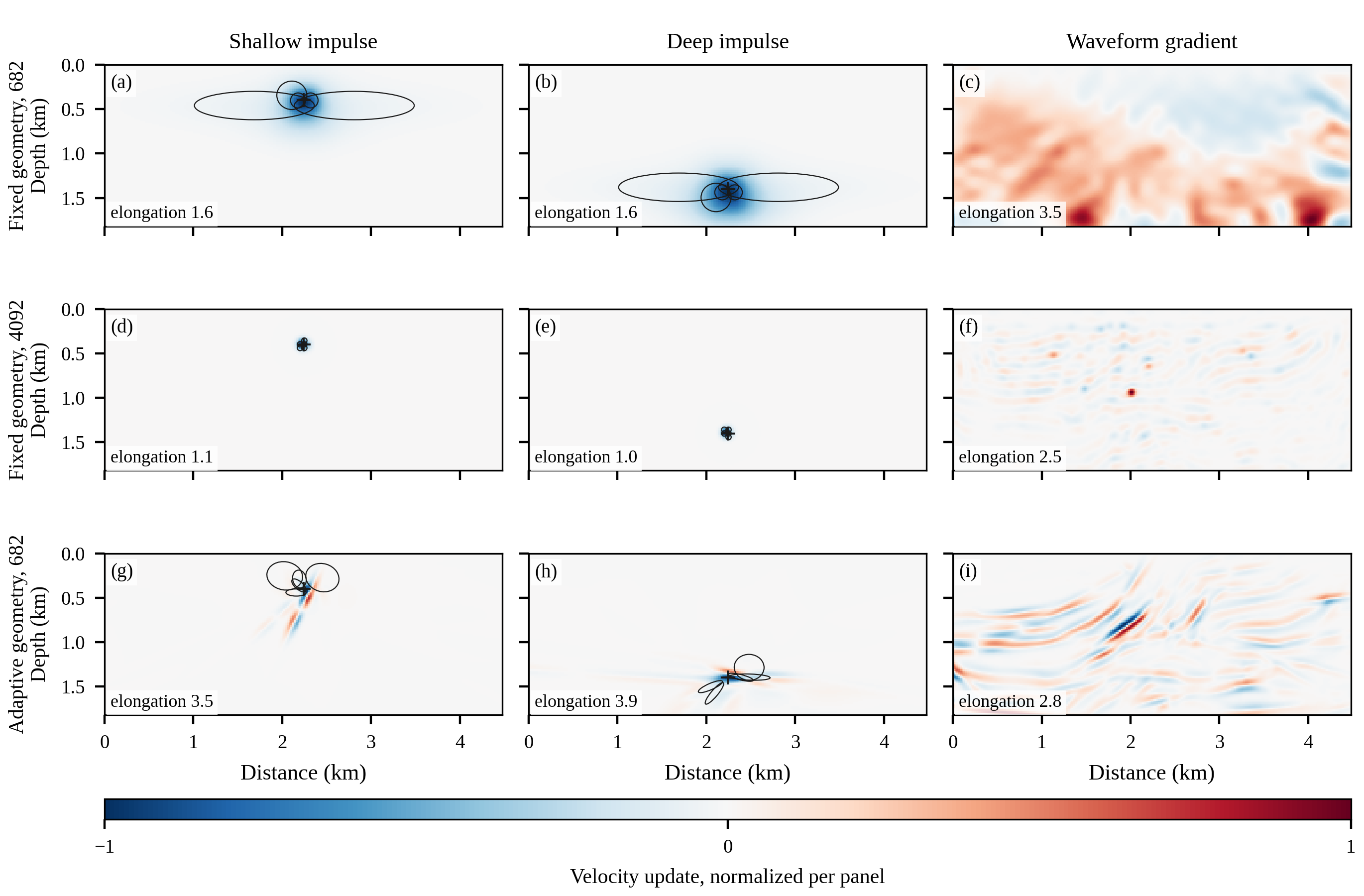}
  \caption{Model-space responses $-\mathbf{P}_\theta \mathbf{g}$ at update 1000 for the representations in Fig.~\ref{fig:marmousi-geometry}(a)--(c). The first two columns show responses to unit impulses at 0.40 and 1.40~km depth, both at a horizontal distance of 2.24~km. The third column shows the response to the waveform gradient evaluated in the model of the coarse fixed run. Crosses mark the impulse locations. From top to bottom, the rows show fixed geometry with 682 primitives, fixed geometry with 4092 primitives, and adaptive geometry with 682 primitives. Each panel is normalized to its maximum absolute value. The ellipses show the one-standard-deviation contours of the five primitives nearest to each impulse. Elongation is the square root of the ratio of the principal second moments of the response energy.}
  \label{fig:update-structure}
\end{figure}

We evaluate the operators at update 1000 for the coarse fixed, dense fixed, and adaptive representations without refinement (Fig.~\ref{fig:marmousi-geometry}(a)--(c)). We apply the same inputs to all three operators so that differences in the responses do not arise from differences in the inputs. Two unit impulses at depths of 0.40 and 1.40~km and a horizontal distance of 2.24~km probe the spatial extent and direction of the response to a local gradient. A common waveform gradient, evaluated in the model of the coarse fixed run at update 1000, shows how these differences carry over to a gradient from the inversion. We repeated the three runs with optimizer states saved. Their final RMSEs differ by less than 0.3~m/s from those in Fig.~\ref{fig:marmousi-geometry}. Because the optimizer scaling can differ between runs, we normalize each panel in Fig.~\ref{fig:update-structure} to its maximum absolute value and compare response shapes rather than amplitudes.

The coarse fixed representation spreads each impulse over a broad region (Fig.~\ref{fig:update-structure}(a) and (b)). The dense fixed representation confines the responses to about one primitive footprint, with elongations of 1.0 to 1.1 (Fig.~\ref{fig:update-structure}(d) and (e)). The adaptive responses have elongations of 3.5 and 3.9 and change sign across their long axes (Fig.~\ref{fig:update-structure}(g) and (h)). They draw on the amplitude, translation, scale, and orientation modes shown in Fig.~\ref{fig:modes}. For the common waveform gradient, the root-mean-square wavenumber falls from 0.159 cycles per cell in the input to 0.036 in the coarse fixed response (Fig.~\ref{fig:update-structure}(c)). The dense fixed and adaptive responses retain more short-wavelength content, with root-mean-square wavenumbers of 0.112 and 0.129 cycles per cell, respectively (Fig.~\ref{fig:update-structure}(f) and (i)). The dense fixed response contains isolated spikes, whereas the adaptive response is more continuous along the reflectors.

To quantify the relation to the layering, we estimate local layer directions from the structure tensor of the true model using Gaussian smoothing scales of 2 and 5 grid cells. At update 1000, 431 of the 682 adaptive primitives have an aspect ratio greater than 1.5, and the median angle between their major axes and the local layer direction is $12.2^\circ$. For the 215 primitives with an aspect ratio greater than 3, the median is $6.6^\circ$, compared with $45^\circ$ for random orientations. We then measure alignment in the response fields themselves. In the input waveform gradient, 53\% of the energy lies within $15^\circ$ of the local layer direction. The corresponding proportions in the coarse fixed, dense fixed, and adaptive responses are 24\%, 59\%, and 69\%, respectively. The adaptive representation thus strengthens the alignment already present in the input gradient.

The measured alignment characterizes the response to the current gradient rather than the full AdamW update. To check the first-order mapping, we replay one AdamW step and use the actual parameter increment $\Delta\theta$. The prediction $\mathbf{B}_\theta\Delta\theta$ matches the resulting velocity change to within 0.2\% for the adaptive representation and exactly for the fixed representations, which are linear in the amplitudes. This agreement tests the linearization given the full parameter step. By contrast, eq.~(\ref{eq:update-operator}) uses only the current gradient. In the replayed step, the contribution from past gradients is six to nine times as large in model space as that from the current gradient. Their magnitudes for the adaptive run are 86 and 9.3~m/s, respectively. The geometric constraint projection contributes a change of 90~m/s in magnitude, and the net velocity change has a magnitude of 38~m/s. Within this diagnostic, the more continuous response along the layers is consistent with the better velocity recovery of the adaptive representation at the same parameter count (Fig.~\ref{fig:marmousi-geometry}(b) and (c)).

\subsection{Comparison with conventional and neural parameterizations}
\label{sec:comparison}

We compare Gauss-FWI on the Marmousi model with four other parameterizations under the same acquisition geometry and starting model. Traditional grid-based FWI updates the velocity of every cell directly (20\,700 parameters) without regularization. Multiscale grid FWI uses the same grid-based parameterization as grid FWI and applies frequency continuation \citep{bunks1995}. We refer to this baseline as Multiscale FWI. It low-pass filters the observed and predicted data at 5, 8, and 12~Hz in three successive stages of 1250 updates, using a cosine taper up to 1.5 times each cutoff. The final 1250 updates use the full band. The B-spline parameterization has 2891 coefficients, and IFWI represents the model with a coordinate-based network of 986\,113 weights.

\begin{figure}[tbp]
  \centering
  \includegraphics[width=0.85\textwidth,height=0.82\textheight,keepaspectratio]{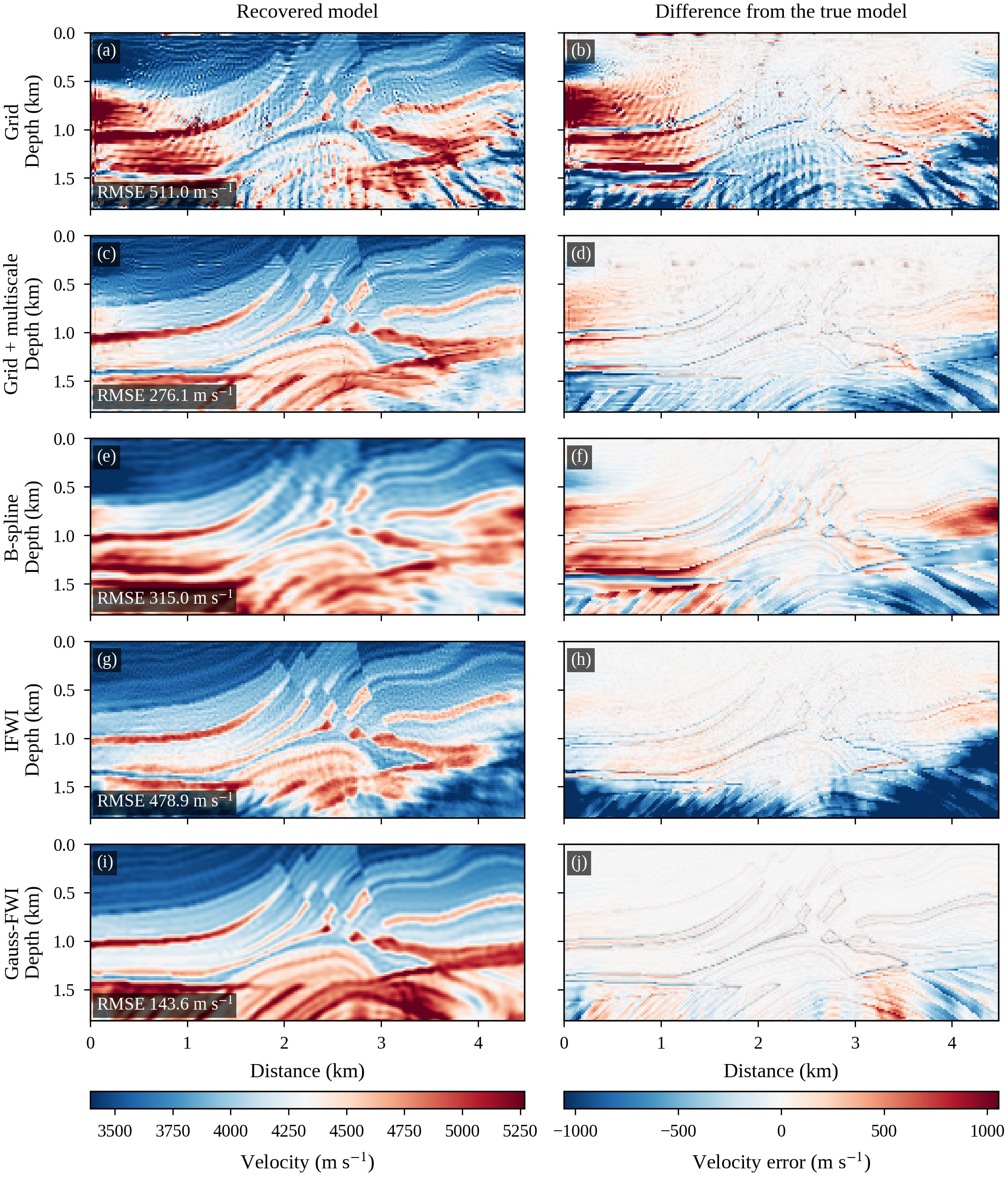}
  \caption{Marmousi velocity models (left) and their differences from the true model (right) obtained with Grid, Grid + multiscale, B-spline, IFWI, and Gauss-FWI after 5000 updates. The RMSE of each recovered model is shown in the left column.}
  \label{fig:marmousi-baselines}
\end{figure}

All inversions in this comparison follow the 5000-update budget specified above. Gauss-FWI uses the same configuration as the full Gauss-FWI run in Section~\ref{sec:ablation} and ends with 3358 primitives and 20\,148 trainable parameters. Fig.~\ref{fig:marmousi-baselines} shows the recovered velocity models and the differences obtained by subtracting the true velocity from each recovered model.

The grid inversion without frequency continuation has an RMSE of 511.0~m/s (Fig.~\ref{fig:marmousi-baselines}(a) and (b)). Its model contains high-wavenumber striping throughout, velocities up to about 1000~m/s too high between 0.5 and 1.0~km depth near the left boundary, and velocities that are too low in the deepest part of the model. Frequency continuation removes most of these artefacts. Multiscale FWI recovers the layered structure with an RMSE of 276.1~m/s (Fig.~\ref{fig:marmousi-baselines}(c) and (d)), and its remaining errors are concentrated below about 1.4~km in the lower left and lower right corners. The B-spline model captures the large-scale geometry but smooths out the thin layers (Fig.~\ref{fig:marmousi-baselines}(e) and (f)). A high-velocity error remains near the right boundary at about 0.8~km depth, and errors persist along the deep layers on the left. Its RMSE is 315.0~m/s.

IFWI recovers the shallow and intermediate layers sharply (Fig.~\ref{fig:marmousi-baselines}(g)), and its difference from the true model is small above about 1.3~km (Fig.~\ref{fig:marmousi-baselines}(h)). Below this depth, however, the model is too slow over most of the section and most strongly at the lower right, which raises its RMSE to 478.9~m/s. The deep and lateral parts of the model are less well constrained by the finite acquisition geometry. There the observations may permit different model updates that fit the waveforms similarly, so the result depends on the parameterization.

Gauss-FWI has the lowest RMSE, 143.6~m/s (Fig.~\ref{fig:marmousi-baselines}(i)). Its difference from the true model consists mainly of thin errors along interfaces, with a few larger errors near the bottom of the model at the lower left and at about 3.5~km distance (Fig.~\ref{fig:marmousi-baselines}(j)). Below 1.0~km depth, its relative velocity error is 4.22\%, compared with 8.30\% for Multiscale FWI, 8.98\% for B-spline, 14.12\% for Grid, and 15.03\% for IFWI. Learning the centres, scales, and orientations changes the spatial structure of the model updates. Refinement adds local degrees of freedom. Combining geometry adaptation with refinement helps Gauss-FWI recover finer structure while retaining the spatial coupling within each primitive.

A closer data fit does not imply a more accurate model. The median normalized objective over the final 200 updates is $6.0 \times 10^{-5}$ for Multiscale FWI and $8.6 \times 10^{-5}$ for Gauss-FWI, so the grid inversion with frequency continuation fits the training waveforms slightly better, yet its deep relative error is almost twice as large. IFWI, with a median objective of $1.3 \times 10^{-4}$, fits the data better than B-spline ($4.9 \times 10^{-3}$) and Grid ($4.6 \times 10^{-4}$) but has the largest deep error.

\subsection{Performance under challenging data conditions}
\label{sec:stress}

\begin{figure}[tbp]
  \centering
  \includegraphics[width=\textwidth,height=0.82\textheight,keepaspectratio]{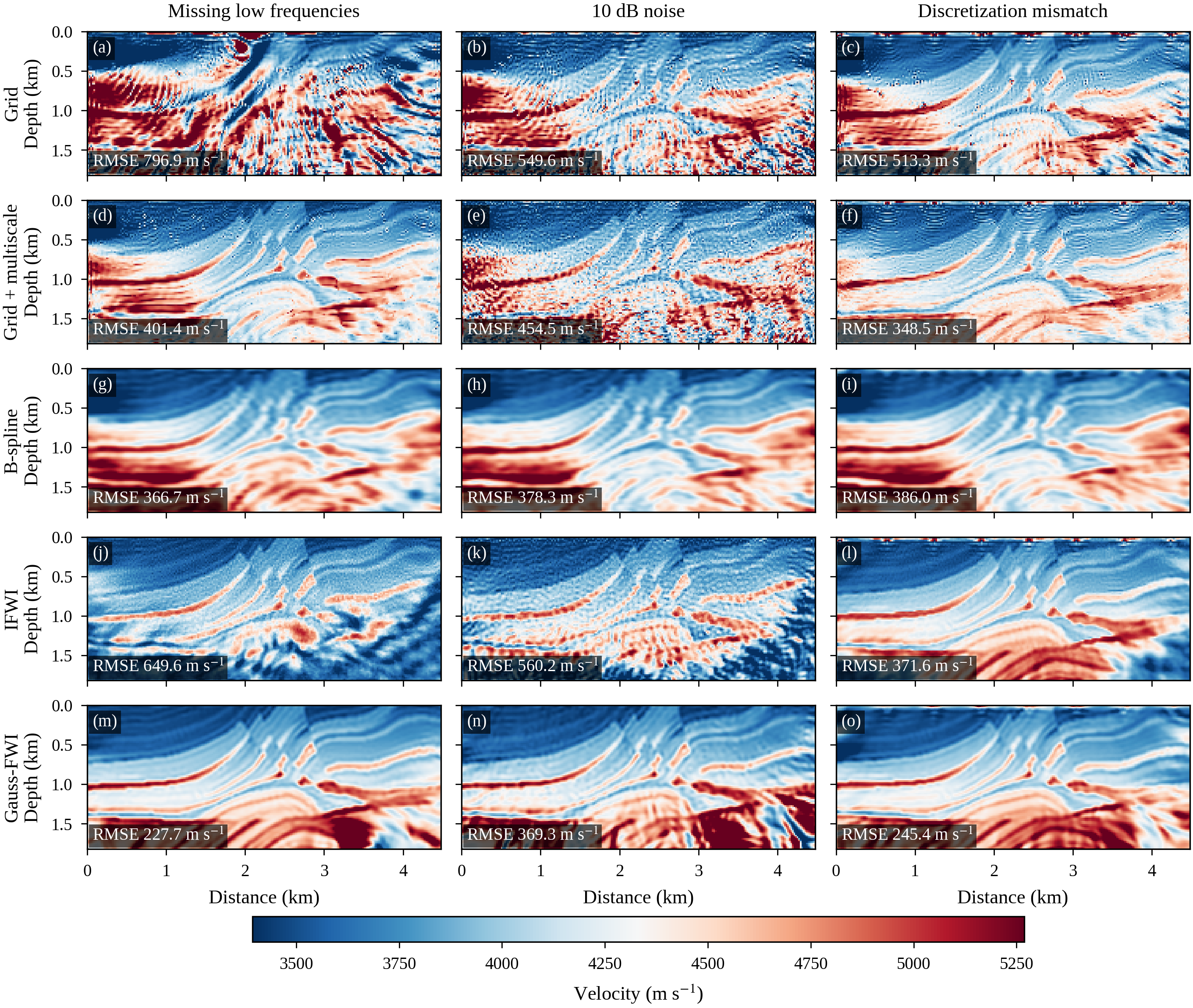}
  \caption{Marmousi velocity models recovered under challenging data conditions. The columns show results for missing low frequencies, 10~dB noise, and discretization mismatch. The rows show Grid, Grid + multiscale, B-spline, IFWI, and Gauss-FWI. The RMSE of each recovered model is shown in the corresponding panel.}
  \label{fig:marmousi-stress}
\end{figure}

We further assess the five parameterizations under three challenging data conditions. The experiments consider missing low-frequency content, additive noise, and a mismatch between the discretizations used to generate the observations and perform the inversion. All methods use the same starting model and acquisition geometry within each experiment. Fig.~\ref{fig:marmousi-stress} presents the recovered velocity models after 5000 updates.

\subsubsection{Missing low-frequency data}

To examine the effect of missing low-frequency information, a high-pass filter is applied to both the observed and predicted waveforms. The filter suppresses frequencies below 8~Hz and has a cosine transition between 8 and 12~Hz. Reduced low-frequency content can make the long-wavelength velocity structure more difficult to recover and increase the sensitivity of waveform matching to the starting model. Because the data contain no energy below 8~Hz, Multiscale FWI uses low-pass stages at 12 and 15~Hz followed by the full band.

The results are shown in the first column of Fig.~\ref{fig:marmousi-stress}. The grid inversion fails without low-frequency data. High-wavenumber artefacts cover the model, a strong anomaly forms near the surface at about 2~km distance, and the RMSE is 796.9~m/s (Fig.~\ref{fig:marmousi-stress}(a)). Frequency continuation recovers the shallow layers but leaves strong high-velocity errors on the left below about 0.7~km depth (Fig.~\ref{fig:marmousi-stress}(d), 401.4~m/s). The B-spline result is smooth and retains the broad layered geometry (Fig.~\ref{fig:marmousi-stress}(g), 366.7~m/s). IFWI recovers several shallow and intermediate-depth interfaces, including parts of the central dipping structure, but its deep model is patchy and too slow, particularly towards the lateral boundaries (Fig.~\ref{fig:marmousi-stress}(j), 649.6~m/s).

Gauss-FWI recovers the main layered and curved structures and has the lowest RMSE, 227.7~m/s (Fig.~\ref{fig:marmousi-stress}(m)). It avoids the broad low-velocity regions of IFWI at depth but overestimates deep velocities between about 3 and 3.7~km distance. Because Gauss-FWI adjusts existing primitives and introduces finer ones during the inversion, the available model updates change as the velocity estimate develops, which may reduce the reliance on the low frequencies for building the long-wavelength structure.

\subsubsection{Noisy data}

Seismic observations are inevitably contaminated by noise, which can introduce spurious components into the model updates and degrade velocity recovery. To examine the sensitivity of the inversion to such disturbances, we add noise to the synthetic observations to obtain an overall signal-to-noise ratio of 10~dB. The same noisy data are used for all five inversions. The resulting models are shown in the second column of Fig.~\ref{fig:marmousi-stress}.

Both grid inversions contain speckled high-wavenumber noise throughout the model (Fig.~\ref{fig:marmousi-stress}(b) and (e)), with RMSEs of 549.6 and 454.5~m/s. Frequency continuation improves the background but does not suppress the noise imprint of the full-band stage. The smooth B-spline model retains the main layered geometry (Fig.~\ref{fig:marmousi-stress}(h), 378.3~m/s). IFWI recovers several major layers but contains oscillatory variations and a deep model that is too slow (Fig.~\ref{fig:marmousi-stress}(k), 560.2~m/s).

Gauss-FWI retains clean shallow and intermediate-depth layers without the speckle of the grid results (Fig.~\ref{fig:marmousi-stress}(n)). Its deep right part, below about 1.3~km and beyond 3~km distance, nevertheless contains strong high- and low-velocity anomalies. Its RMSE of 369.3~m/s is the lowest, but only slightly below the 378.3~m/s of B-spline. The smooth B-spline basis avoids large deep anomalies, whereas the adaptive Gaussian representation preserves shallow detail.

\subsubsection{Discretization mismatch}

In practical FWI, the wavefields predicted by a numerical propagator may differ from the observations even when the velocity model is accurate. Differences in spatial grid spacing, finite-difference order, and the discrete representation of the source can introduce phase and amplitude discrepancies in the simulated data. These discrepancies may affect the model updates as the inversion attempts to fit the observations. To investigate their influence, the reference observations are generated on a 10~m grid using eighth-order spatial differences, while the inversions use the 20~m grid and fourth-order spatial differences described above. The same reference observations, starting model, and acquisition geometry are used for all five methods.

The results are shown in the third column of Fig.~\ref{fig:marmousi-stress}. The grid inversions recover the main interfaces but show localized anomalies along the top of the model near the source positions and errors at depth (Fig.~\ref{fig:marmousi-stress}(c) and (f), 513.3 and 348.5~m/s). Such near-source variations can arise when the inversion attempts to reduce waveform differences associated with mismatched numerical propagation or source representation. IFWI and Gauss-FWI show weaker near-surface anomalies, and the smooth B-spline model shows the least (Fig.~\ref{fig:marmousi-stress}(i), 386.0~m/s). IFWI recovers the shallow and intermediate layers well, but its deep model is again too slow towards the right boundary (Fig.~\ref{fig:marmousi-stress}(l), 371.6~m/s).

Gauss-FWI recovers the principal layered structure with the lowest RMSE, 245.4~m/s (Fig.~\ref{fig:marmousi-stress}(o)), while local errors remain near the surface and in the deep model. Its primitives organize the updates over local regions, and their learnable geometry lets the extent and direction of those updates adapt during the inversion, which may help limit extreme local variations under the tested mismatch.

\subsection{Sensitivity to the starting model}
\label{sec:starting-model}

\begin{figure}[tbp]
  \centering
  \includegraphics[width=\textwidth,height=0.82\textheight,keepaspectratio]{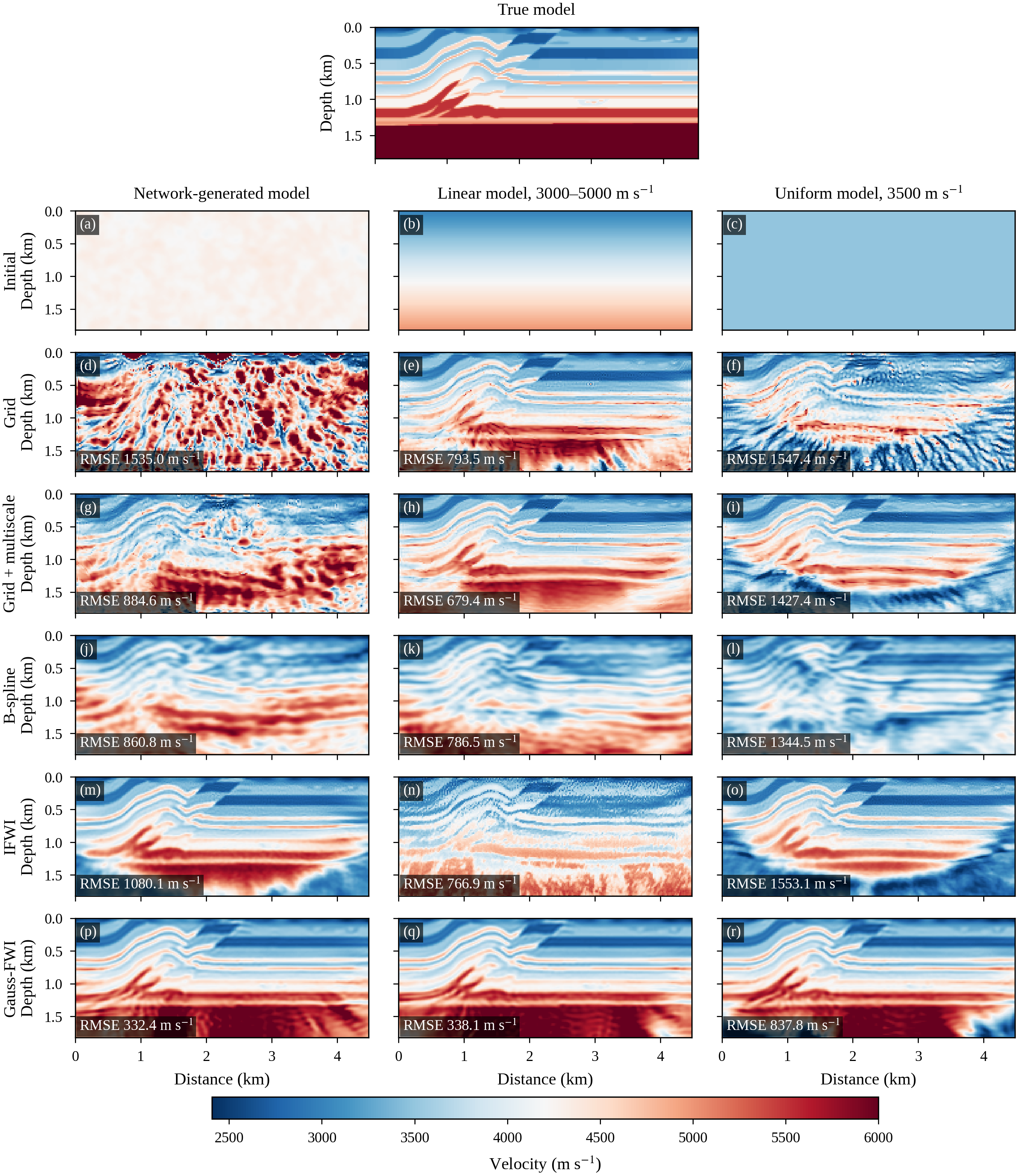}
  \caption{Overthrust velocity models recovered from three starting models. The true model is shown at the top. The columns show the common network-generated model, a model increasing linearly with depth from 3000 to 5000~m/s, and a uniform model of 3500~m/s. The first row shows the starting models, and the following rows show the results of Grid, Grid + multiscale, B-spline, IFWI, and Gauss-FWI after 5000 updates. The RMSE of each recovered model is shown in its panel.}
  \label{fig:overthrust-initials}
\end{figure}

The starting velocity model affects the wave paths and arrival times predicted at the first iteration of FWI. Large initial phase discrepancies can cause the waveform objective to align the wrong waveform cycles and drive the inversion towards an inaccurate model. Even when the arrivals align, the starting model affects the sensitivity of the simulated wavefield to model updates and can influence which structures are recovered. We compare several starting models to examine how these effects vary with the initial velocity distribution.

To examine starting-model sensitivity beyond Marmousi, we use an Overthrust-derived section with folded and faulted layers above a deep high-velocity region (Fig.~\ref{fig:initial-waveform}(d)). The two examples have the same grid, acquisition geometry, and source wavelet. The common starting model in Fig.~\ref{fig:initial-waveform}(b) and (e) was generated by a randomly initialized network and scaled using a vertical velocity profile from Marmousi. Its initial velocity RMSE is 526.5~m/s for Marmousi and 1157.3~m/s for Overthrust, and the relative waveform errors over all eight shots are 63.00\% and 97.72\%, respectively. The selected traces in Fig.~\ref{fig:initial-waveform}(c) and (f) illustrate the initial waveform mismatch for the two models.

For Overthrust, we compare the common network-generated model, a velocity model increasing linearly with depth from 3000 to 5000~m/s, and a uniform model of 3500~m/s (Fig.~\ref{fig:overthrust-initials}(a)--(c)). Their initial velocity RMSEs are 1157.3, 811.6, and 1513.8~m/s, respectively.

We compute first-arrival traveltimes by fast marching for all $8 \times 224$ source--receiver pairs in the true, starting, and recovered models. Table~\ref{tab:traveltime} reports the mean absolute traveltime difference from the true model and the percentage of pairs for which the absolute difference exceeds half the period at the source wavelet's peak frequency. The threshold is approximately 33~ms for the 15~Hz Marmousi and Overthrust data. Exceeding it indicates a potential risk of cycle skipping. The two BP salt cases in the table are discussed in Section~\ref{sec:bp-salt}. For the common network-generated start, the mean absolute first-arrival difference is 72.6~ms for Marmousi and 96.5~ms for Overthrust, and the proportions above the threshold are 73\% and 87\%, respectively. The first-arrival statistics and velocity RMSE rank the three Overthrust starts differently. The linear start has the lowest velocity RMSE, but the uniform start has a smaller mean first-arrival difference (18.2 versus 36.9~ms) and fewer pairs above the threshold (17\% versus 45\%). The network-generated start has a lower velocity RMSE than the uniform start but the largest first-arrival difference.

\begin{table}[tbp]
\centering
\caption{First-arrival traveltime difference between each model and the true model. Each entry gives the mean absolute difference over the $8 \times 224$ source--receiver pairs in ms and, in parentheses, the percentage of pairs whose difference exceeds half a period of the peak frequency (33~ms at 15~Hz for Marmousi and Overthrust, 62.5~ms at 8~Hz for BP salt). Traveltimes are computed by fast marching on the 20~m inversion grid. The smallest mean difference among the recovered models is shown in bold.}
\label{tab:traveltime}
\small
\begin{tabular}{@{}lrrrrrr@{}}
\toprule
Model, starting model & Starting model & Grid & Multiscale & B-spline & IFWI & Gauss-FWI \\
\midrule
Marmousi, network & 72.6 (73) & 17.9 (27) & 1.5 (0) & 2.4 (0) & 0.6 (0) & \textbf{0.3 (0)} \\
Overthrust, network & 96.5 (87) & 103.4 (81) & 18.8 (23) & 10.1 (5) & \textbf{0.9 (0)} & 1.6 (0) \\
Overthrust, linear & 36.9 (45) & \textbf{0.8 (0)} & 1.1 (0) & 5.0 (0) & 5.8 ($<$1) & 1.2 (0) \\
Overthrust, uniform & 18.2 (17) & 11.9 (7) & 2.2 (0) & 5.8 (2) & \textbf{0.9 (0)} & 2.4 (0) \\
BP salt, smoothed & 28.1 (10) & 21.1 (8) & \textbf{3.2 (0)} & 35.4 (30) & 3.5 (0) & 12.1 (0) \\
BP salt, linear & 127.9 (52) & 21.1 (2) & 19.3 (0) & 63.1 (44) & 69.5 (45) & \textbf{10.8 (0)} \\
\bottomrule
\end{tabular}
\end{table}

From the network-generated start, the grid inversion produces strong, blotchy velocity fluctuations that obscure the layered structure (Fig.~\ref{fig:overthrust-initials}(d), 1535.0~m/s). Frequency continuation recovers part of the shallow folds, but the deeper model remains noisy (Fig.~\ref{fig:overthrust-initials}(g), 884.6~m/s). B-spline recovers the broad structure with blurred interfaces (Fig.~\ref{fig:overthrust-initials}(j), 860.8~m/s). IFWI recovers the shallow folds sharply, but its deep layers are too slow towards the lower corners (Fig.~\ref{fig:overthrust-initials}(m), 1080.1~m/s). Gauss-FWI produces clear shallow layers and the most complete deep high-velocity background (Fig.~\ref{fig:overthrust-initials}(p)). Its RMSE of 332.4~m/s is similar to that from the linear start. The share of first-arrival differences above the threshold is 87\% before inversion. After inversion, it is 81\% for Grid, 23\% for Multiscale FWI, and 0\% for Gauss-FWI (Table~\ref{tab:traveltime}).

In the linear starting model, all five methods recover parts of the principal folded structure but differ in interface definition and in the deep high-velocity region. Grid and Multiscale FWI produce clean shallow images (Fig.~\ref{fig:overthrust-initials}(e) and (h)), but the deep high-velocity background is incomplete or smeared, with RMSEs of 793.5 and 679.4~m/s. B-spline reproduces the broad geometry with blurred interfaces (Fig.~\ref{fig:overthrust-initials}(k), 786.5~m/s). IFWI recovers the folds, but its shallow layers are noisy, and its deep region is patchy (Fig.~\ref{fig:overthrust-initials}(n), 766.9~m/s). Gauss-FWI resolves the shallow interfaces and recovers a continuous deep high-velocity background, with a low-velocity deviation at the lower right corner (Fig.~\ref{fig:overthrust-initials}(q)). Its RMSE of 338.1~m/s is less than half that of any other method from this start.

The uniform initial model contains no velocity increase with depth, and every method has a larger RMSE than from the linear start. The grid inversion recovers shallow layers, but its deep region is too slow and crossed by radial streaks (Fig.~\ref{fig:overthrust-initials}(f), 1547.4~m/s). Multiscale FWI and IFWI recover the shallow folds and the upper part of the deep high-velocity layers inside a bowl-shaped region, but outside it, towards the lateral and lower boundaries, the velocity remains far too low (Fig.~\ref{fig:overthrust-initials}(i) and (o), 1427.4 and 1553.1~m/s). B-spline gives a diffuse model in which the deep high-velocity region is substantially underestimated (Fig.~\ref{fig:overthrust-initials}(l), 1344.5~m/s). Gauss-FWI recovers the deep high-velocity background between about 1.8 and 3.5~km distance and retains the main shallow interfaces, but large low-velocity areas persist at the lower left beneath the fold and at the lower right (Fig.~\ref{fig:overthrust-initials}(r)). Its RMSE increases from 338.1 to 837.8~m/s, so its recovery also deteriorates substantially when the starting background changes from the linear to the uniform model, although it remains the lowest for this start.

Gauss-FWI has the lowest velocity RMSE for all three Overthrust starts, although its recovery deteriorates from the uniform model. First-arrival agreement does not necessarily indicate accurate velocity recovery. From the uniform start, IFWI achieves a mean absolute first-arrival difference of 0.9~ms (Table~\ref{tab:traveltime}), yet its velocity RMSE is 1553.1~m/s. In this acquisition, first arrivals mainly sample the upper part of the model, so the traveltime statistics complement the velocity errors and the recovered structures.

\begin{figure}[tbp]
  \centering
  \includegraphics[width=0.78\textwidth,height=0.80\textheight,keepaspectratio]{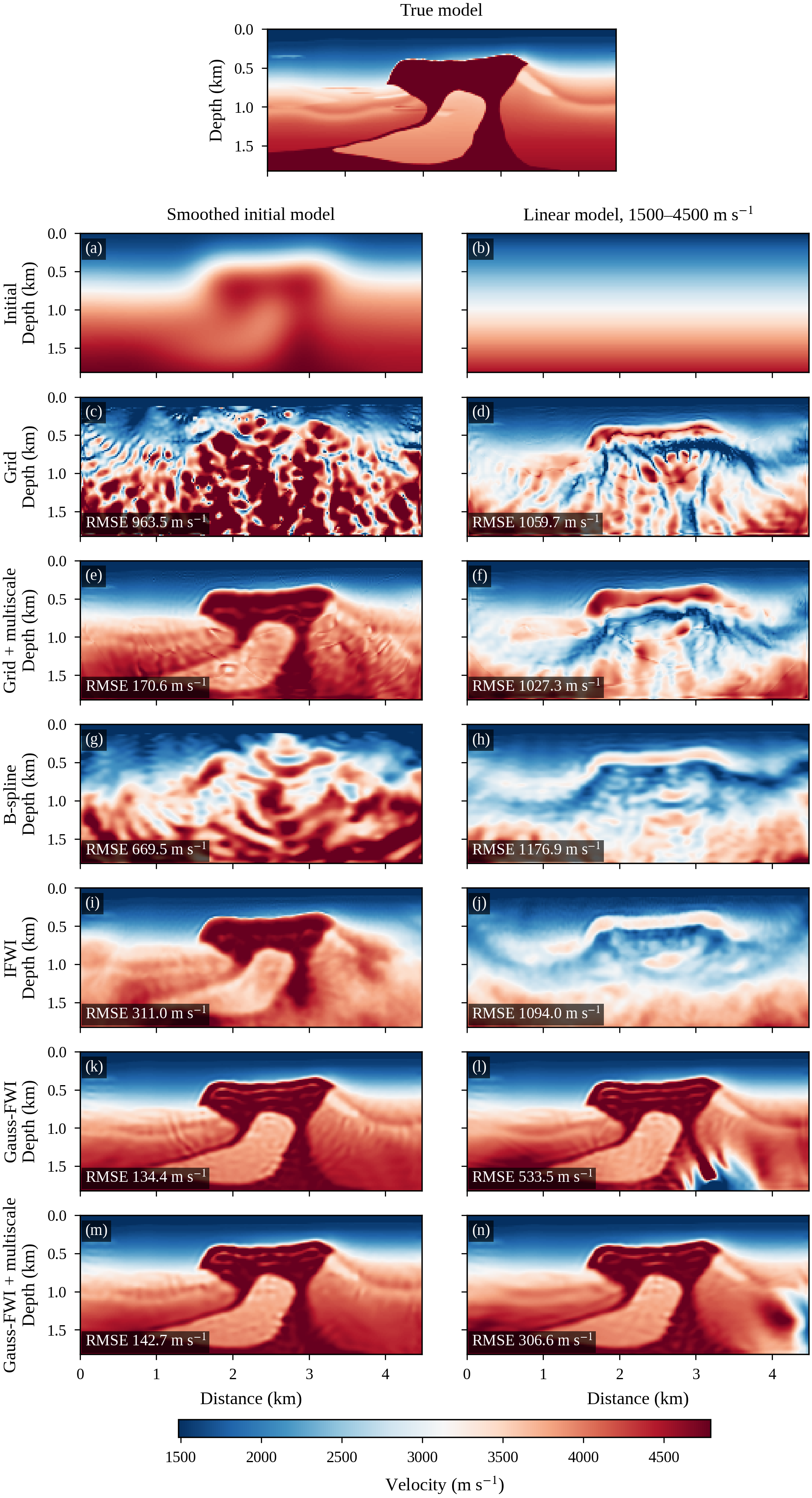}
  \caption{BP salt velocity models recovered from two starting models: (a) a smoothed version of the true model and (b) a model increasing linearly with depth from 1500 to 4500~m/s. The true model is shown at the top. The following rows show the results of Grid, Grid + multiscale, B-spline, IFWI, Gauss-FWI and Gauss-FWI + multiscale after 5000 updates. The water layer is held at its true velocity in all inversions. The RMSE of each recovered model is shown in its panel.}
  \label{fig:bp-salt}
\end{figure}

\subsection{The BP salt model}
\label{sec:bp-salt}

Salt bodies pose a different difficulty. Their strong velocity contrast makes the top and base easy to misplace when the starting model does not contain the salt. We use a section of the 2004 BP velocity benchmark \citep{billette2005} from 26 to 46~km distance over the full depth, compressed anisotropically to the same $92 \times 225$ grid of 20~m cells, so that its geometry is not at the original physical scale. The data are generated with an 8~Hz Ricker wavelet with a peak time of 0.1875~s and the same acquisition geometry as above. The water layer, defined by the cells with velocities at or below 1500~m/s in the true model, is held at its true velocity in all inversions. We start from a smoothed version of the true model (velocity RMSE 348.2~m/s) and from a model increasing linearly with depth from 1500 to 4500~m/s (870.5~m/s). Multiscale FWI uses low-pass stages at 3, 5, and 8~Hz. We also compare Gauss-FWI without frequency continuation against Gauss-FWI with the same frequency continuation schedule (Fig.~\ref{fig:bp-salt}).

From the smoothed start, the grid inversion without continuation fails (Fig.~\ref{fig:bp-salt}(c), 963.5~m/s), and B-spline does not resolve the salt body (Fig.~\ref{fig:bp-salt}(g), 363.5~m/s). Multiscale FWI, IFWI, and Gauss-FWI recover the salt with its top, flanks, and stem (Fig.~\ref{fig:bp-salt}(e), (i), and (k)), with RMSEs of 170.6, 311.0, and 134.4~m/s. IFWI blurs the sediments around the salt. The frequency continuation strategy does not improve Gauss-FWI from this start (Fig.~\ref{fig:bp-salt}(m), 142.7~m/s). The linear start contains no salt, and Grid, Multiscale FWI, B-spline, and IFWI recover at most the top of the salt at about 0.4--0.5~km depth without building the body beneath it (Fig.~\ref{fig:bp-salt}(d), (f), (h), and (j)), with RMSEs between 1027.3 and 1176.9~m/s. Gauss-FWI recovers the salt body and the dipping sediments from this start (Fig.~\ref{fig:bp-salt}(l)), but it produces a spurious low-velocity zone with a narrow high-velocity finger below about 1.3~km between 3.5 and 4.2~km distance, and its RMSE is 533.5~m/s. Gauss-FWI with frequency continuation confines this error to a localized anomaly near the lower right boundary and lowers the RMSE to 306.6~m/s (Fig.~\ref{fig:bp-salt}(n)). Frequency continuation therefore complements the Gaussian representation when the starting model lacks the main structure, whereas it does not rescue the grid inversion from the linear start.

The BP salt results show why first-arrival agreement alone is insufficient for evaluating deep recovery. From the linear start, Multiscale FWI reduces the mean absolute first-arrival difference from 127.9 to 19.3~ms and the share above the 62.5~ms half-period threshold from 52\% to 0\% (Table~\ref{tab:traveltime}). Yet the recovered model does not reconstruct the salt body below its top (Fig.~\ref{fig:bp-salt}(f)). Gauss-FWI also brings the share above the threshold to 0\% and recovers more of the salt body even without frequency continuation (Fig.~\ref{fig:bp-salt}(l)). Across the six starts in Table~\ref{tab:traveltime}, Gauss-FWI is the only method for which the reported share above the threshold is 0\% in every case. Adding continuation to Gauss-FWI improves salt recovery from the linear start but not from the smoothed start. The benefit of continuation in this example therefore depends on both the starting model and the parameterization.

\section{Discussion}
\label{sec:discussion}

\subsection{Roles of geometry, refinement, and the vertical bound}

Gauss-FWI changes the space of model updates in two complementary ways. Adaptive geometry reshapes the existing update patterns, and refinement adds new ones. With the same 4092 trainable parameters, the adaptive representation reduced the Marmousi RMSE from 441.5 to 254.5~m/s relative to the dense fixed dictionary, and refinement then reduced it to 143.6~m/s (Section~\ref{sec:ablation}). The update-structure analysis of Section~\ref{sec:update-structure} explains the first effect. In a fixed dictionary, the operator $\mathbf{P}_\theta$ changes only through the optimizer metric, so the spatial pattern of every update is set when the dictionary is chosen. With adaptive geometry, the columns of $\mathbf{B}_\theta$ follow the model. After 1000 updates, the elongated primitives were aligned with the layers to within a median of 12.2$^\circ$, and the resulting updates were continuous along the reflectors and richer in short wavelengths than those of the dense fixed dictionary. The representation thus learns from the waveform data alone where to couple the model and in which direction, a choice that a fixed basis or a hand-designed regularizer has to make in advance.

Refinement adds capacity where it is needed. Because the children start with zero amplitude, each refinement event leaves the model and the data fit unchanged and only enlarges the set of update directions, and the gradient-energy score places the new directions where the data still drive the model. On Marmousi, this sharpened the thin layers and reduced the deep overshoot that remained with adaptive geometry alone (Fig.~\ref{fig:marmousi-geometry}(d)). The capacity of the representation therefore grows with the inversion, coarse where the model is already explained and fine where the residual structure lies.

The vertical bound $\sigma_{z,\max}$ keeps each primitive within a limited depth range. Without it, a primitive could extend from the well-illuminated shallow part into the weakly constrained deep part, and the strong shallow gradient would then drive deep velocity changes through the shared amplitude and geometry. The bound also limits the vertical extent of the children, which inherit half the scales of their parent. It keeps the coupling introduced by the primitives along the layers rather than across depth levels, which is consistent with the small deep errors of Gauss-FWI on Marmousi and Overthrust.

\subsection{The prior implied by the Gaussian representation}

Because the data constrain the model incompletely, especially in its deep and lateral parts, the representation acts as an implicit prior that selects among models with similar data fit. Multiscale FWI fits the Marmousi waveforms slightly better than Gauss-FWI, with median objectives of $6.0 \times 10^{-5}$ and $8.6 \times 10^{-5}$, yet its deep relative error is almost twice as large (Section~\ref{sec:comparison}). Where the acquisition provides little constraint, the parameterization determines the result.

The Gaussian representation favours a particular kind of model. It starts from $v_0$ with all amplitudes at zero, and weight decay pulls the amplitudes back towards zero, so a perturbation persists only where the data support it. Each perturbation is a sum of smooth, localized, and elongated bumps whose orientation follows the local structure. This suits layered and folded sediments, in which the velocity varies slowly along the layers and rapidly across them, as in the Marmousi and Overthrust examples. The prior differs from conventional regularization in two respects. It is not stationary, because the geometry of the primitives, and with it the direction and length of the coupling, is learned during the inversion. It is also expressed through the parameterization rather than through a penalty term, which is the sense in which neural reparameterizations act as priors \citep{wu2025}. The operator $\mathbf{P}_\theta$ is a metric-weighted counterpart of the standard model-space neural tangent kernel ($\mathbf{J}\mathbf{J}^\top$) in \citet{chen2026wave}, rather than their wave-based neural tangent kernel, which maps through the forward-modelling Jacobian into the data space; both model-space operators describe how parameter updates are mapped into the velocity domain. In Gauss-FWI, this kernel is given explicitly by the positions, scales, and orientations of the primitives and can be read directly from them.

The same prior is less suited to sharp and laterally extensive contrasts. A velocity step across a salt boundary has to be built from many small primitives near the minimum scale, so the salt body draws heavily on the refinement budget. From the linear BP start, Gauss-FWI built the salt body together with a spurious low-velocity zone below it, and frequency continuation removed most of this error (Fig.~\ref{fig:bp-salt}(l) and (n)). The error therefore came from the inversion path rather than from a lack of expressiveness.

The prior also shapes the model in regions that the data constrain poorly. With 10~dB noise, Gauss-FWI kept clean shallow layers but formed strong anomalies in the deep right part of the model, where the smooth B-spline basis stayed closer to the background. The representation does not distinguish gradient components caused by signal from those caused by noise. Where the illumination is weak, only the amplitude decay restrains the growth of deep primitives, and the refinement score adds capacity wherever the gradient energy is high. The flexibility that recovers detail in well-illuminated regions thus also admits artefacts in poorly illuminated ones.

\subsection{Limitations and outlook}
\label{sec:outlook}

The examples in this study are two-dimensional constant-density acoustic synthetics, and all Gauss-FWI settings were shared across the Marmousi, Overthrust, and BP salt models. Extending the method to three dimensions is mainly a question of evaluating the primitives efficiently. The cost per update is dominated by the wave-equation solves, and the dense evaluation of $K$ primitives on $N$ grid nodes, which costs $O(KN)$ operations, is small on the grids used here. In three dimensions, it can be replaced by truncated footprints and tile-based evaluation, as in Gaussian splatting for rendering \citep{kerbl2023}.

For field data, the behaviour under noise points to two extensions. Normalizing the refinement score by an illumination estimate would let new capacity follow data sensitivity rather than raw gradient energy, and primitives whose amplitudes remain small could be pruned or merged. Coupling neighbouring primitives would further stabilize poorly illuminated regions. In anchor-based Gaussian splatting, the attributes of nearby Gaussians are decoded from shared features on a sparse set of anchors \citep{lu2024}. A similar construction would let well-constrained regions inform poorly constrained ones, combining the explicit geometry of Gauss-FWI with the parameter sharing of implicit representations.

The explicit form of the representation also invites further extensions. Because the representation only changes the mapping from the physical gradient to the parameters (eq.~(\ref{eq:parameter-gradient})), it combines directly with frequency continuation, as in the BP salt example, and with misfit functions designed against cycle skipping, such as traveltime or optimal-transport measures \citep{hu2026,yang2018}. For multiparameter or elastic inversion, each primitive can carry one amplitude per physical parameter while sharing its centre, scales, and orientation, which couples the parameters through a common structural geometry. The initial dictionary need not be a uniform lattice. Its orientations can be set from the structure tensor of a migrated image or of the first gradient. Feed-forward Gaussian splatting methods, which predict the primitives of a scene from images in a single network pass \citep{charatan2024}, suggest an amortized variant in which a trained network proposes the primitives and Gauss-FWI refines them against the waveform data, combining a learned starting configuration with physics-based refinement.

\section{Conclusions}

We have presented Gauss-FWI, a full-waveform inversion framework in which the velocity perturbation is represented as a sum of anisotropic Gaussian primitives with learnable amplitudes, centres, scales, and orientations. The wave equation is solved on a fixed grid, and the analytic derivatives of the primitives propagate the adjoint-state gradient to all parameters, allowing the method to integrate directly into standard FWI workflows. Three properties distinguish Gauss-FWI from grid, spline, and neural parameterizations. First, the geometry of the basis is learned from the waveform data: primitives move onto interfaces, stretch along layers, and rotate with the geological dip, which turns the waveform gradient into continuous, layer-parallel model updates and yields far more accurate reconstructions than a denser fixed dictionary with the same parameter budget. Second, the capacity of the representation grows progressively where the data require it: gradient-guided refinement introduces half-scale, zero-amplitude primitives that enlarge the local update space without perturbing the current velocity model. Third, the representation is compact and interpretable: each primitive carries an explicit position, extent, and orientation, and the final Marmousi model uses about 20\,000 parameters, fifty times fewer than implicit FWI.

In two-dimensional acoustic experiments on the Marmousi, Overthrust, and BP salt models, Gauss-FWI achieved lower velocity errors than grid, multiscale, B-spline, and implicit neural parameterizations. It remained robust under missing low frequencies, additive noise, and propagator discretization mismatch, and recovered the principal subsurface structures from inaccurate starting models where the other parameterizations failed. In particular, it reconstructed the BP salt body from a linear starting model without prior salt geometry, and combining the representation with frequency continuation further suppressed deep artefacts. Adaptive Gaussian primitives thus provide an effective and interpretable model parameterization for regularizing FWI, with three-dimensional, multiparameter, and field-data inversions as natural next steps.

\section*{Data Availability}
The Marmousi, SEG/EAGE Overthrust, and 2004 BP velocity models used in this study are publicly available. The synthetic data were generated with Deepwave \citep{richardson2026}, which is openly available on Zenodo. 

\bibliographystyle{abbrvnat}
\bibliography{references}

\begin{thebibliography}{41}
\providecommand{\natexlab}[1]{#1}
\providecommand{\url}[1]{\texttt{#1}}
\expandafter\ifx\csname urlstyle\endcsname\relax
  \providecommand{\doi}[1]{doi: #1}\else
  \providecommand{\doi}{doi: \begingroup \urlstyle{rm}\Url}\fi

\bibitem[Audia et~al.(2025)Audia, Feizi, Zwicker, and Manocha]{audia2025}
S.~Audia, S.~Feizi, M.~Zwicker, and D.~Manocha.
\newblock How learnable grids recover fine detail in low dimensions: a neural
  tangent kernel analysis of multigrid parametric encodings.
\newblock In \emph{Proc. International Conference on Learning Representations
  (ICLR)}, 2025.
\newblock arXiv:2504.13412.

\bibitem[Barnier et~al.(2019)Barnier, Biondi, and Clapp]{barnier2019}
G.~Barnier, E.~Biondi, and R.~G. Clapp.
\newblock Waveform inversion by model reduction using spline interpolation.
\newblock In \emph{SEG Technical Program Expanded Abstracts}, pages 1400--1404.
  SEG, 2019.
\newblock \doi{10.1190/segam2019-3216866.1}.

\bibitem[Barnier et~al.(2023)Barnier, Biondi, Clapp, and Biondi]{barnier2023}
G.~Barnier, E.~Biondi, R.~G. Clapp, and B.~Biondi.
\newblock Full-waveform inversion by model extension: theory, design, and
  optimization.
\newblock \emph{Geophysics}, 88\penalty0 (5):\penalty0 R579--R607, 2023.
\newblock \doi{10.1190/geo2022-0350.1}.

\bibitem[Billette and Brandsberg-Dahl(2005)]{billette2005}
F.~J. Billette and S.~Brandsberg-Dahl.
\newblock The {2004 BP} velocity benchmark.
\newblock In \emph{67th EAGE Conference \& Exhibition, Extended Abstracts},
  page B035. EAGE, 2005.
\newblock \doi{10.3997/2214-4609-pdb.1.b035}.

\bibitem[Bodin and Sambridge(2009)]{bodin2009}
T.~Bodin and M.~Sambridge.
\newblock Seismic tomography with the reversible jump algorithm.
\newblock \emph{Geophys. J. Int.}, 178\penalty0 (3):\penalty0 1411--1436, 2009.
\newblock \doi{10.1111/j.1365-246X.2009.04226.x}.

\bibitem[Borcea et~al.(2023)Borcea, Garnier, Mamonov, and
  Zimmerling]{borcea2023}
L.~Borcea, J.~Garnier, A.~V. Mamonov, and J.~Zimmerling.
\newblock Waveform inversion via reduced order modeling.
\newblock \emph{Geophysics}, 88\penalty0 (2):\penalty0 R175--R191, 2023.
\newblock \doi{10.1190/geo2022-0070.1}.

\bibitem[Bozda{\u{g}} et~al.(2011)Bozda{\u{g}}, Trampert, and
  Tromp]{bozdag2011}
E.~Bozda{\u{g}}, J.~Trampert, and J.~Tromp.
\newblock Misfit functions for full waveform inversion based on instantaneous
  phase and envelope measurements.
\newblock \emph{Geophys. J. Int.}, 185\penalty0 (2):\penalty0 845--870, 2011.
\newblock \doi{10.1111/j.1365-246X.2011.04970.x}.

\bibitem[Bunks et~al.(1995)Bunks, Saleck, Zaleski, and Chavent]{bunks1995}
C.~Bunks, F.~M. Saleck, S.~Zaleski, and G.~Chavent.
\newblock Multiscale seismic waveform inversion.
\newblock \emph{Geophysics}, 60\penalty0 (5):\penalty0 1457--1473, 1995.
\newblock \doi{10.1190/1.1443880}.

\bibitem[Charatan et~al.(2024)Charatan, Li, Tagliasacchi, and
  Sitzmann]{charatan2024}
D.~Charatan, S.~L. Li, A.~Tagliasacchi, and V.~Sitzmann.
\newblock {pixelSplat}: {3D} {Gaussian} splats from image pairs for scalable
  generalizable {3D} reconstruction.
\newblock In \emph{Proc. IEEE/CVF Conference on Computer Vision and Pattern
  Recognition (CVPR)}, pages 19457--19467. IEEE, 2024.
\newblock \doi{10.1109/CVPR52733.2024.01840}.

\bibitem[Chen et~al.(2020)Chen, Albertin, Nihei, Kaplan, Hoversten, Zhang, and
  Bevc]{chen2020swi}
J.~Chen, U.~K. Albertin, K.~Nihei, S.~Kaplan, M.~Hoversten, L.~Zhang, and
  D.~Bevc.
\newblock Stochastic full-waveform inversion ({SWI}) of seismic data for {FWI}
  starting models.
\newblock In \emph{SEG Technical Program Expanded Abstracts}, pages 840--844.
  SEG, 2020.
\newblock \doi{10.1190/segam2020-3418905.1}.

\bibitem[Chen et~al.(2025)Chen, Wu, Li, and Luo]{chen2025}
R.~Chen, B.~Wu, M.~Li, and Y.~Luo.
\newblock Full-waveform inversion with velocity model low-rank implicit neural
  representation.
\newblock \emph{IEEE Trans. Geosci. Remote Sens.}, 63:\penalty0 1--16, 2025.
\newblock \doi{10.1109/TGRS.2025.3594184}.

\bibitem[Chen et~al.(2026{\natexlab{a}})Chen, Luo, Wu, and Meng]{chen2026wave}
R.~Chen, Y.~Luo, B.~Wu, and D.~Meng.
\newblock Unveiling the mechanism of continuous representation full-waveform
  inversion: a wave based neural tangent kernel framework.
\newblock In \emph{Proc. International Conference on Learning Representations
  (ICLR)}, 2026{\natexlab{a}}.
\newblock URL \url{https://openreview.net/forum?id=blqYa21WOv}.
\newblock arXiv:2603.22362.

\bibitem[Chen et~al.(2026{\natexlab{b}})Chen, Sun, Zhu, Song, and
  Tan]{chen2026pert}
Y.~Chen, J.~Sun, Z.~Zhu, P.~Song, and J.~Tan.
\newblock Perturbation-based implicit full-waveform inversion with improved
  optimization conditioning.
\newblock \emph{Geophys. J. Int.}, 247\penalty0 (3):\penalty0 ggag403,
  2026{\natexlab{b}}.
\newblock \doi{10.1093/gji/ggag403}.

\bibitem[Du et~al.(2024)Du, Sun, Jia, Wang, and Liu]{du2024}
B.~Du, J.~Sun, A.~Jia, N.~Wang, and H.~Liu.
\newblock Physics-informed robust and implicit full waveform inversion without
  prior and low-frequency information.
\newblock \emph{IEEE Trans. Geosci. Remote Sens.}, 62:\penalty0 1--12, 2024.
\newblock \doi{10.1109/TGRS.2024.3416547}.

\bibitem[Fichtner et~al.(2013)Fichtner, Trampert, Cupillard, Saygin, Taymaz,
  Capdeville, and Villase{\~n}or]{fichtner2013}
A.~Fichtner, J.~Trampert, P.~Cupillard, E.~Saygin, T.~Taymaz, Y.~Capdeville,
  and A.~Villase{\~n}or.
\newblock Multiscale full waveform inversion.
\newblock \emph{Geophys. J. Int.}, 194\penalty0 (1):\penalty0 534--556, 2013.
\newblock \doi{10.1093/gji/ggt118}.

\bibitem[He and Wang(2021)]{he2021}
Q.~He and Y.~Wang.
\newblock Reparameterized full-waveform inversion using deep neural networks.
\newblock \emph{Geophysics}, 86\penalty0 (1):\penalty0 V1--V13, 2021.
\newblock \doi{10.1190/geo2019-0382.1}.

\bibitem[Hu et~al.(2026)Hu, Huang, Xu, and Ba]{hu2026}
Y.~Hu, X.~Huang, Y.~Xu, and J.~Ba.
\newblock Waveform-preserving source-independent wave-equation-based
  local-scale traveltime inversion: application to the {SEG} {Chevron} 2014
  blind test data set.
\newblock \emph{Geophys. J. Int.}, 247\penalty0 (2):\penalty0 ggag357, 2026.
\newblock \doi{10.1093/gji/ggag357}.

\bibitem[Kadu et~al.(2017)Kadu, van Leeuwen, and Mulder]{kadu2017}
A.~Kadu, T.~van Leeuwen, and W.~A. Mulder.
\newblock Salt reconstruction in full-waveform inversion with a parametric
  level-set method.
\newblock \emph{IEEE Trans. Comput. Imaging}, 3\penalty0 (2):\penalty0
  305--315, 2017.
\newblock \doi{10.1109/TCI.2016.2640761}.

\bibitem[Kang et~al.(2025)Kang, Chen, Yang, Li, and Wu]{kang2025}
B.~Kang, R.~Chen, K.~Yang, M.~Li, and B.~Wu.
\newblock Implicit full waveform inversion with adaptive {Fourier} frequency
  bases learning.
\newblock \emph{Geophys. J. Int.}, 244\penalty0 (1):\penalty0 ggaf404, 2025.
\newblock \doi{10.1093/gji/ggaf404}.

\bibitem[Kerbl et~al.(2023)Kerbl, Kopanas, Leimk{\"u}hler, and
  Drettakis]{kerbl2023}
B.~Kerbl, G.~Kopanas, T.~Leimk{\"u}hler, and G.~Drettakis.
\newblock {3D} {Gaussian} splatting for real-time radiance field rendering.
\newblock \emph{ACM Trans. Graph.}, 42\penalty0 (4):\penalty0 139:1--139:14,
  2023.
\newblock \doi{10.1145/3592433}.

\bibitem[Lailly(1983)]{lailly1983}
P.~Lailly.
\newblock The seismic inverse problem as a sequence of before stack migrations.
\newblock In J.~B. Bednar, R.~Redner, E.~Robinson, and A.~Weglein, editors,
  \emph{Conference on Inverse Scattering: Theory and Application}, pages
  206--220. SIAM, 1983.

\bibitem[Loris et~al.(2007)Loris, Nolet, Daubechies, and Dahlen]{loris2007}
I.~Loris, G.~Nolet, I.~Daubechies, and F.~A. Dahlen.
\newblock Tomographic inversion using {$\ell_1$}-norm regularization of wavelet
  coefficients.
\newblock \emph{Geophys. J. Int.}, 170\penalty0 (1):\penalty0 359--370, 2007.
\newblock \doi{10.1111/j.1365-246X.2007.03409.x}.

\bibitem[Lu et~al.(2024)Lu, Yu, Xu, Xiangli, Wang, Lin, and Dai]{lu2024}
T.~Lu, M.~Yu, L.~Xu, Y.~Xiangli, L.~Wang, D.~Lin, and B.~Dai.
\newblock {Scaffold-GS}: structured {3D} {Gaussians} for view-adaptive
  rendering.
\newblock In \emph{Proc. IEEE/CVF Conference on Computer Vision and Pattern
  Recognition (CVPR)}, pages 20654--20664. IEEE, 2024.
\newblock \doi{10.1109/CVPR52733.2024.01952}.

\bibitem[M{\"u}ller et~al.(2022)M{\"u}ller, Evans, Schied, and
  Keller]{muller2022}
T.~M{\"u}ller, A.~Evans, C.~Schied, and A.~Keller.
\newblock Instant neural graphics primitives with a multiresolution hash
  encoding.
\newblock \emph{ACM Trans. Graph.}, 41\penalty0 (4):\penalty0 102:1--102:15,
  2022.
\newblock \doi{10.1145/3528223.3530127}.

\bibitem[Plessix(2006)]{plessix2006}
R.-E. Plessix.
\newblock A review of the adjoint-state method for computing the gradient of a
  functional with geophysical applications.
\newblock \emph{Geophys. J. Int.}, 167\penalty0 (2):\penalty0 495--503, 2006.
\newblock \doi{10.1111/j.1365-246X.2006.02978.x}.

\bibitem[Pratt et~al.(1998)Pratt, Shin, and Hicks]{pratt1998}
R.~G. Pratt, C.~Shin, and G.~J. Hicks.
\newblock {Gauss--Newton} and full {Newton} methods in frequency--space seismic
  waveform inversion.
\newblock \emph{Geophys. J. Int.}, 133\penalty0 (2):\penalty0 341--362, 1998.
\newblock \doi{10.1046/j.1365-246X.1998.00498.x}.

\bibitem[Rahaman et~al.(2019)Rahaman, Baratin, Arpit, Draxler, Lin, Hamprecht,
  Bengio, and Courville]{rahaman2019}
N.~Rahaman, A.~Baratin, D.~Arpit, F.~Draxler, M.~Lin, F.~Hamprecht, Y.~Bengio,
  and A.~Courville.
\newblock On the spectral bias of neural networks.
\newblock In \emph{Proc. 36th International Conference on Machine Learning
  (ICML)}, volume~97, pages 5301--5310. PMLR, 2019.
\newblock URL \url{https://proceedings.mlr.press/v97/rahaman19a.html}.

\bibitem[Richardson(2026)]{richardson2026}
A.~Richardson.
\newblock {Deepwave}, v0.0.26.
\newblock Zenodo, 2026.
\newblock URL \url{https://doi.org/10.5281/zenodo.18187133}.

\bibitem[Sun et~al.(2003)Sun, Chang, and Sheng]{sun2003}
G.~Sun, Q.~Chang, and P.~Sheng.
\newblock Full waveform inversion with optimal basis functions.
\newblock \emph{Phys. Rev. Lett.}, 90\penalty0 (10):\penalty0 104301, 2003.
\newblock \doi{10.1103/PhysRevLett.90.104301}.

\bibitem[Sun et~al.(2023)Sun, Innanen, Zhang, and Trad]{sun2023}
J.~Sun, K.~Innanen, T.~Zhang, and D.~Trad.
\newblock Implicit seismic full waveform inversion with deep neural
  representation.
\newblock \emph{J. Geophys. Res.: Solid Earth}, 128\penalty0 (3):\penalty0
  e2022JB025964, 2023.
\newblock \doi{10.1029/2022JB025964}.

\bibitem[Tarantola(1984)]{tarantola1984}
A.~Tarantola.
\newblock Inversion of seismic reflection data in the acoustic approximation.
\newblock \emph{Geophysics}, 49\penalty0 (8):\penalty0 1259--1266, 1984.
\newblock \doi{10.1190/1.1441754}.

\bibitem[Versteeg(1994)]{versteeg1994}
R.~Versteeg.
\newblock The {Marmousi} experience: velocity model determination on a
  synthetic complex data set.
\newblock \emph{The Leading Edge}, 13\penalty0 (9):\penalty0 927--936, 1994.
\newblock \doi{10.1190/1.1437051}.

\bibitem[Virieux and Operto(2009)]{virieux2009}
J.~Virieux and S.~Operto.
\newblock An overview of full-waveform inversion in exploration geophysics.
\newblock \emph{Geophysics}, 74\penalty0 (6):\penalty0 WCC1--WCC26, 2009.
\newblock \doi{10.1190/1.3238367}.

\bibitem[Wang and Alkhalifah(2026)]{wang2026}
S.~Wang and T.~Alkhalifah.
\newblock Implicit full waveform inversion imaging.
\newblock \emph{Geophys. J. Int.}, 246\penalty0 (3):\penalty0 ggag277, 2026.
\newblock \doi{10.1093/gji/ggag277}.

\bibitem[Wu et~al.(2014)Wu, Luo, and Wu]{wu2014}
R.-S. Wu, J.~Luo, and B.~Wu.
\newblock Seismic envelope inversion and modulation signal model.
\newblock \emph{Geophysics}, 79\penalty0 (3):\penalty0 WA13--WA24, 2014.
\newblock \doi{10.1190/geo2013-0294.1}.

\bibitem[Wu and Ma(2025)]{wu2025}
Y.~Wu and J.~Ma.
\newblock How does neural network reparametrization improve geophysical
  inversion?
\newblock \emph{J. Geophys. Res.: Machine Learning and Computation}, 2\penalty0
  (2):\penalty0 e2025JH000621, 2025.
\newblock \doi{10.1029/2025JH000621}.

\bibitem[Wu and McMechan(2019)]{wu2019cnn}
Y.~Wu and G.~A. McMechan.
\newblock Parametric convolutional neural network-domain full-waveform
  inversion.
\newblock \emph{Geophysics}, 84\penalty0 (6):\penalty0 R881--R896, 2019.
\newblock \doi{10.1190/geo2018-0224.1}.

\bibitem[Yang et~al.(2018)Yang, Engquist, Sun, and Hamfeldt]{yang2018}
Y.~Yang, B.~Engquist, J.~Sun, and B.~F. Hamfeldt.
\newblock Application of optimal transport and the quadratic {Wasserstein}
  metric to full-waveform inversion.
\newblock \emph{Geophysics}, 83\penalty0 (1):\penalty0 R43--R62, 2018.
\newblock \doi{10.1190/geo2016-0663.1}.

\bibitem[Yuan and Simons(2014)]{yuan2014}
Y.~O. Yuan and F.~J. Simons.
\newblock Multiscale adjoint waveform-difference tomography using wavelets.
\newblock \emph{Geophysics}, 79\penalty0 (3):\penalty0 WA79--WA95, 2014.
\newblock \doi{10.1190/geo2013-0383.1}.

\bibitem[Zhang et~al.(2023)Zhang, Sun, Trad, and Innanen]{zhang2023}
T.~Zhang, J.~Sun, D.~Trad, and K.~Innanen.
\newblock Multilayer perceptron and {Bayesian} neural network-based elastic
  implicit full waveform inversion.
\newblock \emph{IEEE Trans. Geosci. Remote Sens.}, 61:\penalty0 1--16, 2023.
\newblock \doi{10.1109/TGRS.2023.3265657}.

\bibitem[Zhu et~al.(2022)Zhu, Xu, Darve, Biondi, and Beroza]{zhu2022}
W.~Zhu, K.~Xu, E.~Darve, B.~Biondi, and G.~C. Beroza.
\newblock Integrating deep neural networks with full-waveform inversion:
  reparameterization, regularization, and uncertainty quantification.
\newblock \emph{Geophysics}, 87\penalty0 (1):\penalty0 R93--R109, 2022.
\newblock \doi{10.1190/geo2020-0933.1}.

\end{thebibliography}

\end{document}